\documentclass[11pt]{article}
\usepackage{amsmath, amssymb, graphics}

\usepackage[]{cite}
\newcommand{\mathsym}[1]{{}}
\usepackage{verbatim}
\usepackage{amsfonts,euscript,amssymb,stmaryrd,braket}
\usepackage{graphics,tikz}
\usetikzlibrary{shapes.misc,arrows,decorations.markings,patterns}
\usepackage{caption}
\usepackage{subcaption}
\usepackage{slashed}
\definecolor{hyperref}{RGB}{026,028,185}
\usepackage[bookmarks=true,colorlinks=true,linkcolor=hyperref,citecolor=hyperref,urlcolor=hyperref,bookmarksnumbered]{hyperref}
\usepackage{booktabs}
\usepackage{cancel}
\usepackage[normalem]{ulem}
\usepackage{multirow}
\usepackage{tabularx}
\usepackage{longtable}
\usepackage{bbold,bbding}
\usepackage{amsthm} 
\usepackage{esint}
\usepackage{float}
\newcommand{\bal}{\begin{equation}\begin{aligned}}
		\newcommand{\eal}{\end{aligned} \end{equation}}
\newcommand{\ba}{\begin{align}}
\newcommand{\ea}{\end{align}}
\def\id{\protect{{1 \kern-.28em {\rm l}}}}
\usepackage{scalerel}
\usepackage[usestackEOL]{stackengine}
\stackMath

\makeatletter
\renewcommand\section{\@startsection {section}{1}{\z@}%
	{-3.5ex \@plus -1ex \@minus -.2ex}%
	{2.3ex \@plus.2ex}%
	{\normalfont\large\bfseries}}
\renewcommand\subsection{\@startsection{subsection}{2}{\z@}%
	{-3.25ex\@plus -1ex \@minus -.2ex}%
	{1.5ex \@plus .2ex}%
	{\normalfont\normalsize\bfseries}}

\makeatother

\numberwithin{equation}{section}

\usepackage[utf8]{inputenc}
\usepackage{epsfig}
\usepackage{graphicx}
\usepackage[multiple]{footmisc}
\usepackage{amssymb,amsmath}
\usepackage{fancybox,framed,tikz}
\usetikzlibrary{decorations.pathmorphing,patterns}
\usepackage{dsfont}
\usepackage{mathtools}
\usepackage{braket}
\usepackage{slashed}
\usepackage{rotating}
\usepackage{bbold,amsfonts}
\usepackage{multirow}
\usepackage{verbatim}
\usepackage{upgreek}

\tikzset{cross/.style={cross out, draw=black, minimum size=2*(#1-\pgflinewidth), inner sep=0pt, outer sep=0pt},
	cross/.default={1pt}}

\newcommand{\be}{\begin{equation}}
	\newcommand{\ee}{\end{equation}}

\newcommand{\Tr}{\textup{Tr}}

\usepackage{color}

\definecolor{mypink1}{rgb}{0.958, 0.188, 0.478}

\tikzset{Witten diagram/.style={execute at begin picture={%
			\draw[blue ,fill=blue!05] circle[radius=\pgfkeysvalueof{/tikz/Witten/radius}];
			\path node (X){\phantom{X}};
		},baseline={(X.base)}},vertex/.style={circle,fill,inner sep=1.414pt,node
		contents={}},
	Witten/.cd,radius/.initial=1.414cm}

\newcommand\blfootnote[1]{%
	\begingroup
	\renewcommand\thefootnote{}\footnote{#1}%
	\addtocounter{footnote}{-1}%
	\endgroup
}

\begin{document}
	\renewcommand{\thefootnote}{\arabic{footnote}}

	\overfullrule=0pt
	\parskip=2pt
	\parindent=12pt
	\headheight=0in \headsep=0in \topmargin=0in \oddsidemargin=0in
	
	\vspace{ -3cm} \thispagestyle{empty} \vspace{-1cm}
	\begin{flushright} 
		\footnotesize
		{HU-EP-22/14-RTG}
		
	\end{flushright}%

	\begin{center}
		\vspace{1.2cm}
		{\Large\bf \mathversion{bold}
			{Notes on $n$-point Witten diagrams in AdS${}_2$}}
		
		\author{ABC\thanks{XYZ} \and DEF\thanks{UVW} \and GHI\thanks{XYZ}}
		\vspace{0.8cm} {
			Gabriel~Bliard\blfootnote{\tt gabriel.bliard.ph@gmail.com}}
		\vskip  0.5cm
		
		\small
		{\em
			Institut f\"ur Physik, Humboldt-Universit\"at zu Berlin and IRIS Adlershof, \\Zum Gro\ss en Windkanal 2, 12489 Berlin, Germany\\    
			
		}
		\normalsize
		
	\end{center}

	\vspace{0.3cm}
	\begin{abstract} 
Witten diagrams provide a perturbative framework for calculations in Anti-de-Sitter space and play an essential role in a variety of holographic computations. In the case of this study in AdS$_2$, the one-dimensional boundary allows for a simple setup, in which we obtain perturbative analytic results for correlators with the residue theorem. This elementary method is used to find all scalar $n$-point contact Witten diagrams for external operators of conformal dimensions $\Delta=1$ and $\Delta=2$, and to determine topological correlators of Yang-Mills in AdS$_2$. Another established method is applied to explicitly compute exchange diagrams and give an example of a Polyakov block in $d=1$. We also check perturbatively a recently proposed multipoint Ward identity with the strong coupling expansion of the six-point function of operators inserted on the 1/2 BPS Wilson line in $\mathcal{N}$=4 SYM.\par \vspace{1cm}
	\end{abstract}
	
	\newpage
	
	\tableofcontents
	
	\section{Discussion}
		\label{sec:introduction}
		
			 The two-dimensional Anti-de-Sitter space AdS$_2$ provides a wonderful playground both to study quantum field theory in curved space and to further our understanding of correlation functions in AdS/CFT. As there are no propagating degrees of freedom for the graviton, the latter is not a usual gauge/gravity correspondence, but rather a rigid holography that has physical settings e.g. in the context of defects in higher-dimensional theories \cite{Giombi:2017cqn,Liendo:2018ukf,Beccaria:2019dws,Bianchi:2020hsz,Barrat:2020vch,Ferrero:2021bsb}, effective and intrinsic theories in AdS$_2$ \cite{Paulos:2016fap, Ouyang:2019xdd,Mazac:2016qev,Ferrero:2019luz,Beccaria:2020qtk,Beccaria:2019dju,Beccaria:2019stp,Beccaria:2019mev,DiPietro:2017vsp,Maldacena:1998uz,Gross:2017vhb,Maldacena:2016hyu}.  Perturbative computations in Anti-de-Sitter space are done through Witten diagrams, whose structure has extensively been studied \cite{Witten:1998qj,DHoker:1999bve,DHoker:1999kzh,DHoker:1999mqo,Freedman:1998bj,Rastelli:2016nze,Zhou:2018sfz,Zhou:2020ptb,Dolan:2000ut,Penedones:2010ue,Fitzpatrick:2011ia}. Yet, the complexity relative to their flat space counterpart is still a roadblock to perturbative analysis and there is still a search for the full equivalent of Feynman rules \cite{Paulos:2011ie}. As such, the most efficient methods to date for perturbative correlators  are through the conformal bootstrap \cite{Bissi:2022mrs,Gopakumar:2016cpb}. However, explicit computations remain a reliable way to make progress in perturbation theory and can provide some insight into assumptions that may simplify the bootstrap process.  First-order four-point correlators with quartic interactions in the strong coupling limit can be written in terms of $D$-functions \cite{DHoker:1999kzh} which are four-point Witten contact diagrams. At higher order, with loops and exchanges which correspond to additional integrated bulk points, some diagrams can be related to contact integrals through differential equations \cite{DHoker:1999mqo,Rastelli:2017udc}. As such, the $n$-point $D$-functions are used beyond the first order and can be seen as a starting point to build `master integrals' for Witten diagrams. AdS$_2$ is a perfect place to look at these integrals as it provides a simple framework with relevance in its own right (e.g. in defect theories) and corresponds to a diagonal limit ($z= \bar{z}$ for four points) of its higher-dimensional counterparts.\par 
			 \vspace{.2cm}
			Boundary correlators in AdS$_2$ enjoy a one-dimensional conformal symmetry. In higher dimensions, along with additional symmetries, this simplifies perturbative computations. However, the structure of AdS$_2$ provides a framework in which another elementary method can be used to compute perturbative quantities; the residue theorem. Using contour integration for one of the AdS$_2$ integrals, the contact diagram in $\lambda_n \phi_{\Delta}^n$ theory for $n$ scalars of low conformal dimension is remarkably simple leading to the results (see Section \ref{sec:Integration} below)
			\begin{align}
				I_{\Delta=1,n}(x_i)&=\frac{\pi}{(2i)^{n-2}}\sum_{i\neq j}\frac{x_{ij}^{n-4}}{\Pi_{k\neq i, j}x_{ik}x_{jk}}\log\left(\frac{x_{ij}}{2i}\right),\\
				I_{\Delta=2,n}(x_i)&=\sum_{j\neq i}\frac{-\pi}{2(2i)^{2n-4}x_{ij}^2}\partial_{x_i}\left( \frac{x_{ij}^{2n-5}}{\prod_{k\neq  i,j }x_{ik}^2x_{jk}^2}\log\frac{x_{ij}}{2i}\right)\\
				&+ \sum_{j\neq i}\partial_{x_j}\frac{-\pi}{(2i)^{2n-2}x_{ij}^2}\partial_{x_i}\left( \frac{x_{ij}^{2n-4}}{\prod_{k\neq i,j}x_{ik}^2x_{jk}^2} \log\frac{x_{ij}}{2i}\right). \nonumber 
			\end{align}
			Above, $I_{\Delta,n}(x_i)$ is the integral corresponding to the Witten contact diagram of $n$ fields $\phi_\Delta$ of conformal dimension $\Delta$ inserted at positions $x_i$, where $x_{ij}=x_i-x_j$ throughout the text. Here, the normalisation of the propagator is set to one. For the canonical normalisation, one can simply multiply by the quantity defined in \eqref{Eq: normalisation C_Delta} for each propagator. These expressions are in terms of the operators' positions and combine naturally into the cross-ratios obtained with the usual conformal transformations (see discussion in Section \ref{subsec: CFT basics} around equation \eqref{Eq: cross ratio} and Appendix \ref{App: list of correlators}).
			This method proves to be even more powerful in some settings where the residue theorem can be used for both of the bulk coordinates. This is the case for the topological three-point correlator of the boundary fields $a$ of the gauge field of pure Yang-Mills in AdS$_2$ presented in \cite{Mezei:2017kmw}, providing an alternative derivation in Section \ref{sec:topological} of 
			\begin{align}
				\langle a^{a}(x_1)a^{b}(x_2)a^{c}(x_3)\rangle  = -\frac{1}{4\pi g_{YM}^2} f^{abc}\textrm{sgn}(x_{12}x_{23}x_{31}),
			\end{align} 
			where $f^{abc}$ are the structure constants of the gauge group of the Yang-Mills theory. 
			The low dimensionality has other advantages, such as having fewer cross-ratios. For four-point functions, for example, the single cross-ratio simplifies the differential equation used to compute exchange diagrams in Section \ref{sec:polyakov}. We thus compute explicitly a Polyakov block corresponding to the sum of four-point exchange Witten diagrams with dimensions $\Delta=1$, which agrees with independently found results\footnote{We thank Pietro Ferrero for sharing some of the unpublished results related to \cite{Ferrero:2019luz}.}
			\begin{align}
				P_{1,1}^{(0)} (\chi)&=\text{Li}_2\left(\frac{\chi}{\chi-1}\right) \log \left(\frac{\chi^2}{(\chi-1)^2}\right)-6 \text{Li}_3\left(\frac{\chi}{\chi-1}\right)-\frac{1}{6} \pi ^2 \log \left(\frac{\chi^2}{(\chi-1)^2}\right)+6 \zeta (3) \nonumber \\
				&+ \frac{\text{Li}_2(1-\chi) \log \left((\chi-1)^2\right)-6 \text{Li}_3(1-\chi)-\frac{1}{6} \pi ^2 \log \left((\chi-1)^2\right)+6 \zeta (3)}{\chi^2}\nonumber \\
				&+\frac{\text{Li}_2(\chi) \log \left(\chi^2\right)-6 \text{Li}_3(\chi)-\frac{1}{6} \pi ^2 \log \left(\chi^2\right)+6 \zeta (3)}{(\chi-1)^2} ,
			\end{align}\par
		where $\chi=\frac{x_{12}x_{34}}{x_{13}x_{24}}$ is the four-point function cross-ratio.
			Natural extensions to this work include deriving position space results of contact diagrams in the case of higher external dimensions  and Polyakov blocks for higher exchanged dimensions. A possible path for this is by extending the one-dimensional Mellin analysis developed in \cite{Bianchi:2021piu} to higher $n$-point functions using results from this study. In the former, Mellin amplitudes for higher $\Delta$ were obtained, so that, in combination with these notes, results for all $(n,\Delta)$ may be achievable. The knowledge of the structure of contact diagrams also sheds light on the computation of higher-point exchange diagrams through the method presented in Section \ref{sec:polyakov}. A combination of these two techniques for contact and exchange diagrams could also, along with multipoint Ward identities and bootstrap methods, provide a path to the computation of the strong coupling, second-order, 6-point correlator of the 1/2-BPS defect in $\mathcal{N}=4$ SYM. In this spirit, an appendix is included providing a perturbative check of the consistency of the multipoint Ward identities conjectured in \cite{Barrat:2021tpn} for the first two strong coupling perturbative orders of the 6-point correlator of insertions on the 1/2 BPS line in $\mathcal{N}=4$ SYM. The higher-order quantities are beyond the scope of this paper and are left for further investigation. \par
			\vspace{.3cm}
			The paper will proceed as follows; after an introduction to the basics, techniques, and notations of CFT$_1$ and AdS$_2$ in Section \ref{sec:Basics}, contour integration will be used in Section \ref{sec:Integration} to derive the AdS$_2$ massless $n$-point contact diagrams which are consistent with the numerical integration and the current literature. This method will also be used in Section \ref{sec:topological} to derive the topological three-point correlator of the boundary field of pure Yang-Mills in AdS$_2$. Finally, known methods will be applied in Section \ref{sec:polyakov} to find the explicit expression of a one-dimensional Polyakov block, which has the correct symmetries, Regge behaviour, and double-discontinuity. Several technical appendices complete the manuscript.
			

	\section{Review of	\texorpdfstring{AdS$_2$/CFT$_1$}{Lg} correlators}
Interacting fields propagating in AdS$_2$ define a non-local one-dimensional conformal field at the boundary.  We go through some basics of one-dimensional conformal theories, dynamics in two-dimensional Anti-de-Sitter space, and set up the notation of this paper. We also review the concepts of Polyakov blocks and the methods used in \cite{Zhou:2020ptb,Rastelli:2017udc} to relate AdS exchange diagrams to contact diagrams in the context of CFT$_1$.
		\label{sec:Basics}

			\subsection{\texorpdfstring{CFT$_1$}{Lg}  basics, techniques, and notation}\label{subsec: CFT basics}
			
			A one-dimensional conformal field theory can be defined by a set of data consisting of the spectrum $\{\Delta\}$ which defines which operators are present in the theory, and the  coefficients $\{c_{\Delta_1 \Delta_2 \Delta_3}\}$ which define the interactions of these operators. Combined with the Operator Product Expansion (OPE), these can be used to reconstruct any correlator of local operators in the theory. The conformal group in $d=1$ is generated by the translation $P$, dilation $D$, and special conformal transformation $K$ which satisfy the conformal algebra and can be parametrised in one dimension by the differential operators
			\begin{align}
				P&= -\partial_x& D&= -x\partial_x-\Delta&	K&=-x^2\partial_x-2\Delta x,
			\end{align}
			when acting on a field of conformal dimension $\Delta$ evaluated at position $x$.
			The consequence of these symmetries on correlators is that the coordinate dependence of the first three $n$-point correlators is fixed
			\begin{align}
				&\langle \phi_\Delta(x)\rangle  = \delta_{\Delta,0}  \\ 
				&\langle \phi_{\Delta_{1}}(x_1)\phi_{\Delta_2}(x_2)\rangle = \frac{\delta_{\Delta_1,\Delta_2}}{x_{12}^{2\Delta}}\\
				&\langle \phi_{\Delta_{1}}(x_1)\phi_{\Delta_2}(x_2)\phi_{\Delta_3}(x_3)\rangle = \frac{c_{\Delta_1\Delta_2\Delta_3}}{(x_{12})^{\Delta_{123}}(x_{13})^{\Delta_{132}}(x_{23})^{\Delta_{231}}}\qquad \quad \Delta_{ijk} = \Delta_i+\Delta_j-\Delta_k .
			\end{align}
			For $n$-point correlators, conformal transformations can be used to reduce the number of independent variables to $n \!-\!3$ conformally invariant cross-ratios
			\begin{align}\label{Eq: cross ratio}
				\chi_{i-1} &= \frac{x_{1i}x_{n-1,n}}{x_{in}x_{1,n-1}}& 0<&\chi_i<1,
			\end{align}
			where $x_{ij} = x_j-x_i$ are real numbers in $d=1$.
			For equal conformal dimensions the correlator is\footnote{Note that the limit
				\begin{align}
					\lim_{\epsilon\rightarrow 0} 	\langle \phi_\Delta(x_1)..\phi_\Delta(x_i)...\epsilon^{-2\Delta}\phi_\Delta(\epsilon^{-1})\rangle  = 	\langle \phi_\Delta(x_1)..\phi_\Delta(x_i)...\tilde{\phi}_\Delta(\infty)\rangle 
				\end{align}
				is well defined. }
			\begin{align}
				\langle \phi_\Delta(x_1)..\phi_\Delta(x_i)...\phi_\Delta(x_n)\rangle = A(x_1,...,x_n)	\langle \phi_\Delta(0)..\phi_\Delta(\chi_i)...\phi_\Delta(1)\tilde{\phi}_\Delta(\infty)\rangle  ,
			\end{align}
			where
			\begin{align}\label{Eq: prefactor}
				A(x_1,...x_n) &=\left(\left(\frac{x_{1n}x_{n-1,n}}{x_{1,n-1}}\right)^{n-2} \prod _{j=1}^{n-1} x_{jn}^{-2}\right)^{\Delta }.
			\end{align}

			\par
			One might wonder why higher-point correlators are of any interest since the theory can be determined by the set $\{\Delta, c_{\Delta_1\Delta_2\Delta_3} \}$. In practice, however, determining such a set is far from trivial, and working out perturbative higher-point correlators gives access to this information through the dynamics of the theory. For example, the four-point correlator has the following operator product expansion
			\begin{align}\label{Eq: four-point corr}
				<\phi_{\Delta}(x_1)\phi_{\Delta}(x_2)\phi_{\Delta}(x_3)\phi_{\Delta}(x_4)> &=\frac{1}{(x_{13}x_{24})^{2\Delta}}f(\chi)\\
				f(\chi)&= \sum_h c^2_{\Delta \Delta h} \chi^{h-2\Delta}{}_2F_1(h,h,2h;\chi),
			\end{align}
			where $\chi=\frac{x_{12}x_{34}}{x_{13}x_{24}}$ is the usual cross-ratio ($\chi_1$ in \eqref{Eq: cross ratio}) for the four-point function in 1d.
			When computing the analytic expression of this four-point correlator at different orders in perturbation theory, for example using Witten diagrams in a holographic setup \cite{Giombi:2017cqn, Bianchi:2020hsz}, these can be equated to the expansion of \eqref{Eq: four-point corr} to find the perturbative CFT data $\{\Delta, c_{\Delta_1\Delta_2\Delta_3} \}$.\par 
			The symmetries of the conformal blocks (similarity under $\chi\rightarrow \frac{\chi}{\chi-1}$), those of the correlator (crossing symmetry under $\chi\rightarrow1-\chi$), and those of the theory (e.g. Ward identities in \cite{Liendo:2018ukf}) can be used to highly constrain the four-point correlators. Complemented by a transcendentality ansatz for the correlators \cite{Alday:2017zzv}, this provides a powerful way to compute perturbative correlators, as was done in \cite{Ferrero:2021bsb,Liendo:2018ukf,Ferrero:2019luz,Lemos:2016xke,Bianchi:2020hsz}. \par 
			The bosonic four-point correlator defined in \eqref{Eq: four-point corr} has symmetries under the permutations of the external operators. Given the ordering of the operators, the variable $z $ is naturally defined on the range $0<z<1$. Following the analysis of \cite{Mazac:2018qmi}, the bosonic symmetry can be used to define the function in \eqref{Eq: bosonic continuation} on the entire real axis.
			\begin{align}\label{Eq: bosonic continuation}
				f(\chi) = \begin{cases*}
					f^{(-)}(\chi)=(1-\chi)^{2\Delta}f^{(0)}\left(\frac{\chi}{\chi-1}\right) &$\chi<0$\\
					f^{(0)}(\chi) & $0<\chi<1$\\
					f^{(+)}(\chi)=	\chi^{2\Delta}f^{(0)} \left(\frac{1}{\chi}\right)& $\chi>1$
				\end{cases*}
			\end{align}
			The resulting function has an explicit symmetry under crossing  
			\begin{align}
				\chi\rightarrow 1-\chi\qquad 0<\chi<1,
			\end{align}
			and braiding
			\begin{align}
				\chi\rightarrow \frac{\chi}{\chi-1}.
			\end{align}
			These two symmetries generate all the crossing symmetries from bosonic permutations.
			In addition, these functions defined on a segment of the real line can be analytically continued outside their region of analyticity. For some functions (for example, those resulting from contact Witten diagrams), the analytic continuation of the function $f^{(0,\pm)}(\chi)$ outside its segment of definition matches the function $f(\chi)$. In this case, we speak of \textit{braiding symmetry}. This is linked to the vanishing of the double-discontinuity, defined in  \cite{Mazac:2018qmi} as 
			\begin{align}\label{Eq: Ddisc}
				dDisc^{+}[\mathcal{G}(\chi)] = \mathcal{G}^{(0)}(\chi)-\frac{\mathcal{G}^{(+)}(\chi+i\epsilon)+\mathcal{G^{(+)}}(\chi-i \epsilon)}{2} \quad 0<\chi<1.
			\end{align}
			Unitarity arguments link this double-discontinuity to the full correlator thanks to the inversion formula \cite{Mazac:2018qmi} and provide a powerful tool constraining the correlators and correspondingly the OPE data.\par
			The OPE expansion is the projection of the correlator on the basis of the conformal blocks. There is, however, another basis that is of some interest in this context; Polyakov blocks. These are defined to be crossing-symmetric, Regge-bounded,\footnote{The Regge limit of a correlator in $d=1$ is controlled by its behaviour at large $|\chi|$. A Regge bounded correlator $g(\chi)$ satisfies $\lim_{\zeta \rightarrow \infty} g(\frac{1}{2}+i \zeta) < C$ where $C $ is a constant. Note that the Identity contribution has a constant contribution in this limit} and to have the same expansion as the conformal blocks (see for example Section 6 of \cite{Mazac:2018qmi}) 
			\begin{align}
				\sum_h c_h G_h(\chi) = \sum_h c_h P_h(\chi).
			\end{align}
			Their existence in $d=1$ was motivated in \cite{Gopakumar:2016wkt,Gopakumar:2016cpb,Polyakov:1974gs} and proven in \cite{Mazac:2018ycv}. 
			Additionally, they have the same double-discontinuity as the conformal blocks and have a double zero at the position of double trace operators
			\begin{align}
				P_{2\Delta+2n} &=0\\
				\partial_nP_{2\Delta+n} &=\delta_{n,0}. 
			\end{align}
			As a consequence, they can be expressed as the sum of Witten exchange diagrams (see the example of perturbative Polyakov blocks in Appendix \ref{App: Polyakov}). Through the computation of the exchange Witten diagrams in position space in one dimension, the explicit form of a Polyakov block is shown below in Section \ref{sec:polyakov}.
			
			
			\subsection{Witten diagrammatics in \texorpdfstring{AdS$_2$}{Lg}}
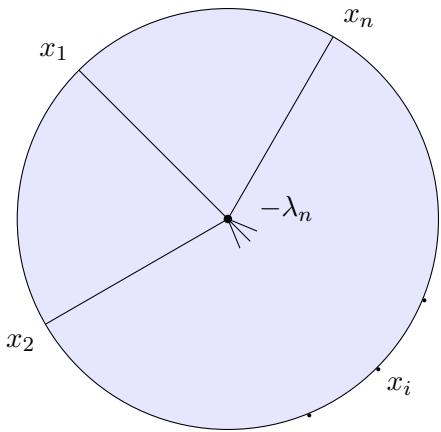
\begin{figure}[h]
				\centering
				\begin{tikzpicture}[scale=.7]
					\def\x{4}
					\filldraw[blue!10!white] (0,0) circle (\x cm);
					\draw[] (0,0) circle (\x cm);
					\node[anchor=west] at (0.1*\x,0.05*\x) {$-\lambda_n$};
					
					\draw[] (0,0) to (-0.707107*\x, 0.707107*\x);
					\node[anchor=south east] at (-0.707107*\x, 0.707107*\x){$x_1$};
					\draw[] (0,0) to  (-0.866025*\x, -0.5*\x);
					\node[anchor=north east] at (-0.866025*\x, -0.5*\x) {$x_2$};
					\draw[] (0,0) to  (0.5*\x,0.866025*\x);
					\node[anchor=south west] at (0.5*\x,0.866025*\x) {$x_n$};
					
					\draw[] (0,0) to  (.3 *0.191342*\x, -.3*0.46194*\x);
					\draw[] (0,0) to  (.3*0.353553*\x, -.3*0.353553*\x);
					\draw[] (0,0) to  (.3*0.46194*\x, -.3*0.191342*\x);
					
					\filldraw[]  (0,0) circle (0.5*\x pt);
					\filldraw[]  (0.191342*\x*2.02, -0.46194*\x*2.02) circle (0.2*\x pt);
					\filldraw[]  (0.353553*\x*2.02, -0.353553*\x*2.02) circle (0.2*\x pt);
					\filldraw[]  (0.46194*\x*2.02, -0.191342*\x*2.02) circle (0.2*\x pt);
					\node[anchor=north west] at (0.353553*\x*2, -0.353553*\x*2){$x_i$};
				\end{tikzpicture}
				\caption{Witten $n$-point contact diagram with a $\lambda_n\phi^n$ interaction and $n$ boundary insertions at positions $\{x_1,...,x_n\}$.}
				\label{Fig:contact-witten}
			\end{figure}

			We consider bulk theories in euclidean AdS$_2$, for which we use the Poincaré metric
			 \begin{align}
				ds^2_{AdS_2} = \frac{dx^2+dz^2}{z^2}.
			\end{align}
		Scalar bosonic fields of mass $m$ are dual to conformal scalars of dimension $\Delta$
			\begin{align}
				m^2 = \Delta(\Delta-1),
			\end{align}
			inserted on the boundary ($z=0$). 
			These fields have a bulk-to-bulk propagator
			\begin{align}\label{Eq: bulk to bulk}
				G^{\Delta}_{ BB}(a,b) &=C_\Delta (2u)^{-1}{}_2F_1(\Delta,\Delta,2\Delta,-2u^{-1})&				u&=\frac{(z_a-z_b)^2+(x_a-x_b)^2}{2z_az_b},
			\end{align}
			which satisfies the AdS$_2$ equation of motion
			\begin{align}
				(\nabla_{AdS}^2-\Delta(\Delta-1)) G^{\Delta}_{ BB}(a,b)=z^2 \delta^{(2)}(a-b)\label{Eq: Equation of motion ads2},
			\end{align}
			and whose normalisation \cite{Freedman:1998tz,Fitzpatrick:2011ia} is
			\begin{align}\label{Eq: normalisation C_Delta}
				C_\Delta = \frac{\Gamma (\Delta )}{2 \sqrt{\pi } \Gamma \left(\Delta +\frac{1}{2}\right)}.
			\end{align}
			The bulk-to-boundary propagator corresponding to the $z\rightarrow 0$ limit of \eqref{Eq: bulk to bulk} is
			\begin{align}\label{Eq: bulk-to-boundary prop}
				K_\Delta(z,x;x_i)&=C_{\Delta}\tilde{K}_\Delta(z,x;x_i)\\
				&=C_{\Delta}\left(\frac{z}{z^2+(x-x_i)^2}\right)^\Delta.
			\end{align}
			Due to the isometries of AdS$_2$, the boundary correlators will be conformal. Given an action, for example the effective worldsheet theory on AdS$_2$ of \cite{Giombi:2017cqn}, boundary correlators are computed via Witten diagrams (e.g. the contact diagram in Figure \ref{Fig:contact-witten}). 
			\par
			Just as in Feynman diagrams, the external legs, propagators and vertices play the same role. The external legs are depicted as points at the boundary and the integral is evaluated over the position of the vertices in the bulk of AdS$_2$.
			For example, the contact diagram depicted in Figure \ref{Fig:contact-witten} corresponds to the integral
			\begin{align}\label{Eq: contact integral 1}
				\lambda_n\tilde{I}_{\Delta,n}(x_1,...,x_n) &= \lambda_n \int \frac{dzdx}{z^2} \prod_{i=1}^{n} K_{\Delta}(z,x;x_i),
			\end{align}
			which will be solved in Section \ref{sec:Integration} for all $n$ and $\Delta=1,2$.\par
			
			The first extension to this class of integrals is to consider multiple bulk integrations. This happens when there are loops and exchanges in the corresponding Witten diagram. For some exchange diagrams, the multiple integrals can be related to \eqref{Eq: contact integral 1} through the action of a differential operator (for more details see Appendix \ref{Sec: relating exchange to contact diagrams}).
			For example, the four-point single-exchange diagram can be found by solving the differential equation
			\begin{align}
				(C_{(34)}^{(2)}-m^2_E)	J(x_1,x_2,x_3,x_4) &=\int \frac{dz_a dx_a}{z_a^2} \Pi_{i=1}^4K_{\Delta_\phi}(z_a,x_a;x_i)\\
				&= \tilde{I}_{\Delta_\phi ,4}(z),
			\end{align}
			where $C^{(2)}_{34}$ is the quadratic Casimir acting on the external legs 3 and 4, $m_E$ is the mass of the exchanged operator, and the full integral is given by
			\begin{align}
				J(x_1...x_4) &= \int \frac{dz_adx_a}{z_a^2} \int \frac{dz_bdx_b}{z_b^2}G^{\Delta_E}_{BB}(a,b)\prod_{i=1}^{2}\left(K_{\Delta_\phi}(z_a,x_a;x_i)\right) \prod_{i=3}^{4}\left( K_{\Delta_\phi}(z_b,x_b;x_i)\right).
			\end{align}
		 	The simple structure of one-dimensional conformal correlators allows us to write the result explicitly in position space for the case $\Delta=\Delta_E=1$, see below in \eqref{Eq: First polyakov block}.
			
			\bigskip
			These quantities have been computed in Mellin space\cite{Mack:2009mi,Penedones:2010ue,Rastelli:2016nze,Fitzpatrick:2011ia,Ferrero:2019luz}, where Witten diagrams have a natural language. In particular, contact diagrams with no derivatives are given by constant truncated Mellin amplitudes and exchange diagrams have poles in Mellin space.\footnote{For an introduction to the subject, a useful resource is \cite{Penedones:2016voo}. }
			However, there are several caveats to these results. Firstly, the Mellin and anti-Mellin transforms are not trivial computations, so knowledge of the Mellin amplitude does not imply that of the position space correlator and vice-versa. Additionally, the generality of such results prevents the use of the simplifications occurring in $d=1$.  Furthermore, when the number of external legs is large enough ($n>d+2$), many spurious Mellin variables do not correspond to a cross-ratio in position space. This is already relevant for the four-point $d=1$ correlator. In one dimension, several attempts were made to use the Mellin transform using the higher-dimensional formalism \cite{Ferrero:2019luz} or developing a one-dimensional formalism \cite{Bianchi:2021piu}. Using as a guide the principle that contact Witten diagrams correspond to constant Mellin amplitudes, the results in Section \ref{sec:Integration} may provide a starting point in generalising the one-dimensional Mellin formalism developed in \cite{Bianchi:2021piu}. There, results for contact diagrams with general external dimensions $\Delta$ were derived so that, in combination with the insights of the present study, results for all $(n,\Delta)$ may be achievable.

	\section{	\texorpdfstring{$n$}{Lg}-point contact diagrams}
		\label{sec:Integration}

			We start by looking at $n$-point correlators of identical scalars with a simple contact interaction. These will serve not only as examples to demonstrate the simplifications occurring in this low-dimensional case, but also as building blocks for the massive contact diagrams, exchange diagrams, and  other cases seen in the following sections. These correlators result from an interaction term $\lambda_n \phi^n$ in the bulk of AdS$_2$ and will be a function of $n-3$ independent cross-ratios due to the symmetry structure of CFT$_1$, or equivalently, the isometry structure of AdS$_2$. These constitute the `master integrals' in AdS$_2$ for contact diagrams used in \cite{Bianchi:2020hsz,Giombi:2017cqn}. The contact diagram is illustrated in Figure \ref{Fig:contact-witten} and can be written as an integration over AdS$_2$ of the $n$ bulk-to-boundary propagators, leading to the connected tree-level correlator
			\begin{align}
				\langle \phi_{\Delta_1}(x_1)...\phi_{\Delta_n}(x_n)\rangle_{conn}^{(1)} &= -\lambda_n \left(\Pi_{i=1}^n C_{\Delta_i} \right)I_{\Delta_1,...,\Delta_n}(x_1,...,x_n),
			\end{align}
			where we define the integral
			\begin{align}\label{Eq: contact integral}
				I_{\Delta_1,...,\Delta_n}(x_1,...,x_n) =\int \frac{dx dz}{z^2}\Pi_{i=1}^n \left(\frac{z}{z^2+(x-x_i)^2}\right)^{\Delta_{i}}.
			\end{align}
			The simplifications of AdS$_2$ can be made explicit by evaluating the $x$-integral first with contour integration. This is especially effective for the massless case ($\Delta=1$) where the integrand of \eqref{Eq: contact integral} only has single poles and the general result -see \eqref{Eq: massless} below- for a massless $n$-point function is derived,
			\begin{align}\label{Eq: solution massless all n}
				\langle \phi_{\Delta=1}(x_1)...\phi(x_n)\rangle = -\lambda_n(C_{\Delta=1})^n	\left \{ 
				\Centerstack{\frac{\pi}{(2i)^{n-2}}\sum_{i> j}\frac{x_{ij}^{n-4}}{\Pi_{k\neq i, j}^n x_{ik} x_{jk}}\log\left(x_{ij}^2\right)\quad   n \quad \textrm{ even}\\
					\quad  \\
					\frac{\pi^2  }{ 2(2i)^{n-3}} \sum_{i>j}^{n}\frac{x_{ij}^{n-4}}{\Pi_{k\neq j\neq i}x_{ik}x_{jk}}\quad n\quad \textrm{ odd}.}
				\right.
			\end{align}
			\subsection{Massless scalar fields}
			\label{subsec:massless}
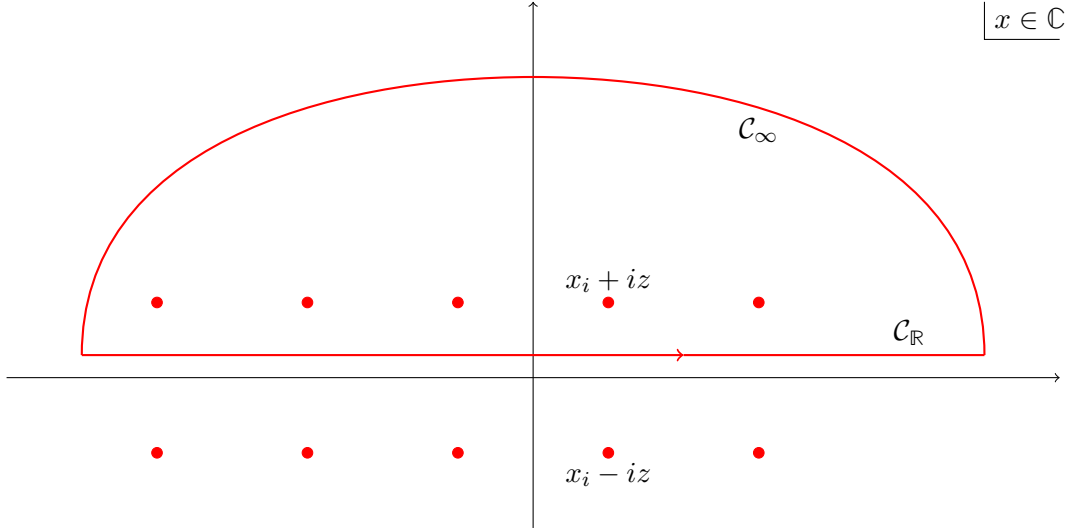
\begin{figure}[H]
				\center
				\begin{tikzpicture}
					\draw[->] (0,-2) to (0,5);
					\draw[->] (-7,0) to (7,0);
					\draw [thick, red, ->] (-6,.3) to (2,.3);
					\draw [thick, red] (2,.3) to (6,.3);
					\filldraw[red] (1,1) circle (2pt);
					\node[anchor= south ] at (1,1)  {$x_i+i z$};
					\filldraw[red] (1,-1) circle (2pt);
					\node[anchor= north ] at (1,-1)  {$x_i-i z$};
					\filldraw[red] (-3,1) circle (2pt);
					\filldraw[red] (-3,-1) circle (2pt);
					\filldraw[red] (-5,1) circle (2pt);
					\filldraw[red] (-5,-1) circle (2pt);
					\filldraw[red] (-1,1) circle (2pt);
					\filldraw[red] (-1,-1) circle (2pt);
					\filldraw[red] (3,1) circle (2pt);
					\filldraw[red] (3,-1) circle (2pt);
					\draw [,thick,red] (-6,.3) to[out=90,in=180] (0,4) to[out=0,in=90](6,.3) ;
					\draw[] (6,5)to(6,4.5) to(7,4.5);
					\node[anchor = south west] at (6,4.5){$x\in \mathbb{C}$};
					\node[anchor= south] at (5,.3) {$\mathcal{C}_{\mathbb{R}}$};
					\node[anchor= south] at (3,3) {$\mathcal{C}_{\infty}$};
				\end{tikzpicture}
				\caption{Contour used for the integral over the variable $x$ parametrising the AdS$_2$ boundary. The contour can be chosen to close in either the upper or lower complex  half-plane since the integrand is appropriately bounded at large $|x|$. }
				\label{Fig: contour integral}
			\end{figure}

			In the case of massless scalar fields, the integral \eqref{Eq: contact integral} reduces to 
			\begin{align}\label{Eq: massless contact integral}
				I_{\Delta=1}(x_1,...,x_n)= \int_0^\infty dz z^{n-2} \int_{-\infty}^{\infty} dx \frac{1}{\Pi_{i=1}^n(z^2+(x-x_i)^2)}.\vspace{.4cm}
			\end{align}
		 The advantage of working in AdS$_2$ is that, since the boundary has only one dimension, the integrated boundary coordinate $x$ can be analytically continued to the complex plane and the integral can be evaluated with the residue theorem. The contribution from the contour around infinity ($\mathcal{C}_\infty$ in Figure \ref{Fig: contour integral}) vanishes since the integrand is appropriately bounded at large $|x|$. 
			
			The integrand in  \eqref{Eq: massless contact integral} has $2n$ poles at
			\begin{align}
				x = x_j\pm i z
			\end{align}
			where $1\leq j\leq n$, with residues
			\begin{align}
				\pm \frac{1}{2iz\Pi_{i\neq j}(x_{ij}^2+2izx_{ij})},
			\end{align}
			which are depicted in Figure \ref{Fig: contour integral}. Since, for $z>0$, the poles in the upper half-plane (UHP) come with a positive sign and those in the lower half-plane (LHP) come with a minus sign, the result is independent of the choice of closing the contour. However, when $z$ is real, these poles will have an additional factor $\textrm{sgn}(z)$. This is because the poles cross the $x\in \mathbb{R}$ axis when $z$ crosses $0$.\par 
			We are thus left with the  integral
			\begin{align}\label{Eq: z-integral}
				I(x_i) =\pi \int_0^\infty dz z^{n-3}\sum_{j=1}^{n}	\frac{1}{\Pi_{1=1,i\neq j}^{n}(x_{ij}^2+2izx_{ij})}.
			\end{align}
			The integrand of \eqref{Eq: z-integral} has a leading large $z$ behaviour 
			\begin{align}
				\pi \sum_{j=1}^{n-1}\frac{z^{-1}}{(2i)^{n-2}}\frac{1}{\Pi_{i=1,i\neq j}^{n}x_{ij}}+O(z^{-2}),
			\end{align}
			which vanishes thanks to the identity
			\begin{align}
				\sum_{j\in J}\frac{1}{\Pi_{i\in J, i\neq j}x_{ij}}=0,
			\end{align}
			so the integral is convergent for $n\geq 3$ as expected. Notice that the integrand of  \eqref{Eq: z-integral} has the same parity as the number of external fields. This leads to a simplification in computing the odd $n$-point functions. 
			\subsubsection*{$n$ odd}
\begin{figure}[H]
				\center
				\begin{tikzpicture}[scale=.5]
					\draw[->] (0,-2) to (0,5);
					\draw[->] (-7,0) to (7,0);
					\draw [thick, red, ->] (-6,.3) to (2,.3);
					\draw [thick, red] (2,.3) to (6,.3);
					\filldraw[red] (1,1) circle (2pt);
					\node[anchor= south ] at (1.5,1)  {$x-i z$};
					\filldraw[red] (1,-1) circle (2pt);
					\node[anchor= north ] at (1.5,-1)  {$x+i z$};
					\filldraw[red] (-1,1) circle (2pt);
					\filldraw[red] (-1,-1) circle (2pt);
					\draw [,thick,red] (-6,.3) to[out=90,in=180] (0,4) to[out=0,in=90](6,.3) ;
					\draw[] (6,5)to(6,4.5) to(7,4.5);
					\node[anchor = south west] at (6,4.5){$x\in \mathbb{C}$};
					\node[anchor= south] at (5,.3) {$\mathcal{C}_{\mathbb{R}}$};
					\node[anchor= south] at (3.5,3.5) {$\mathcal{C}_{\infty}$};
				\end{tikzpicture}
				\begin{tikzpicture}[scale=.5]
					\draw[->] (0,-2) to (0,5);
					\draw[->] (-7,0) to (7,0);
					\draw [thick, red, ->] (-6,.3) to (2,.3);
					\draw [thick, red] (2,.3) to (6,.3);
					\filldraw[red] (1,1) circle (2pt);
					\node[anchor= south ] at (1.5,1)  {$x+i z$};
					\filldraw[red] (1,-1) circle (2pt);
					\node[anchor= north ] at (1.5,-1)  {$x-i z$};
					\filldraw[red] (-1,1) circle (2pt);
					\filldraw[red] (-1,-1) circle (2pt);
					\draw [,thick,red] (-6,.3) to[out=90,in=180] (0,4) to[out=0,in=90](6,.3) ;
					\draw[] (6,5)to(6,4.5) to(7,4.5);
					\node[anchor = south west] at (6,4.5){$x\in \mathbb{C}$};
					\node[anchor= south] at (5,.3) {$\mathcal{C}_{\mathbb{R}}$};
					\node[anchor= south] at (3.5,3.5) {$\mathcal{C}_{\infty}$};
				\end{tikzpicture}
				\caption{The contour here is closed in the UHP (the same analysis holds for the LHP closing). However, since $z$ is now defined on the entire real line the pole enclosed in the given contour depends on the sign of $z$. The case where $z<0$ (Left) will have a residue of opposite sign when compared to that where $z>0$ (Right).}
				\label{Fig: contour integral 2}
			\end{figure}
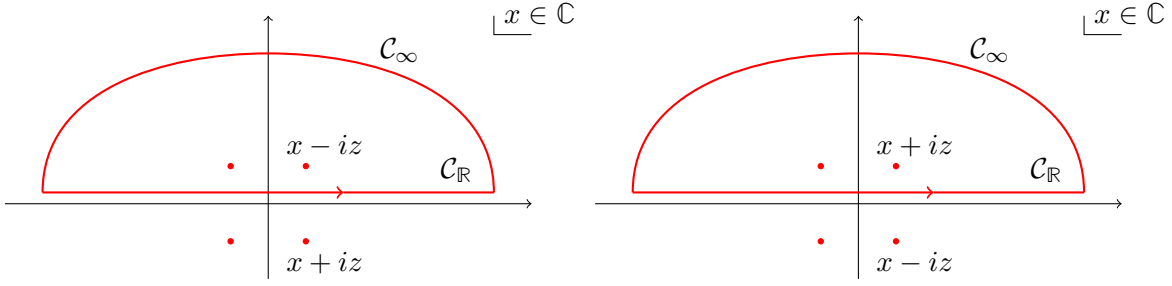

			The massless odd-$n$ case can be solved  with contour integration for both the $x$ and the $z$ coordinates.\footnote{This is true for any convergent integral with odd $\sum_i \Delta_i$. The resulting correlator will thus be a rational polynomial in the cross-ratios though it may not have a simple form.} Since both the integrand and the residue of the pole in the $x$ coordinates are antisymmetric under $z\rightarrow -z$, we can extend the region of integration of $z$ to the entire real line. This is best seen in the trivial example of the conformal massless three-point function
			\begin{align}
				\lim_{\Lambda\rightarrow \infty} \Lambda^{2}I(0,1,\Lambda) = \int_0^\infty \frac{dz}{z^2}\int_{-\infty}^{\infty}dx \frac{z^3}{(z^2+x^2)(z^2+(x-1)^2)}.
			\end{align}
			Since the integrand is antisymmetric under $z\rightarrow -z$, we need to compensate for the sign change when extending the range of the integral over $z$
			\begin{align}\label{Eq 3point sign z}
				\frac{1}{2}\int_{-\infty}^\infty \frac{dz\textrm{sgn}(z)}{z^2}\int_{-\infty}^{\infty}dx \frac{z^3}{(z^2+x^2)(z^2+(x-1)^2)}.
			\end{align}
			When considering the $x$-contour integral \eqref{Eq 3point sign z}, we are now faced with two situations for the contour integral, the first ($z<0$) is depicted on the left of Figure \ref{Fig: contour integral 2}, the second ($z>0$) is depicted on the right.
			The sign of the pole included in the contour cancels the sign contribution from equation \eqref{Eq 3point sign z}. The range of $z$ can be extended after having done the $x$-integral to obtain the same conclusion. In so doing, one obtains
			\begin{align}
				\pi \int_{-\infty}^\infty dz \frac{1 }{4 z^2+1} = \frac{\pi^2}{2}.
			\end{align}
			This reasoning holds for a general $n$-point function whose integrand is antisymmetric under $z\rightarrow -z$.  We will see in one of the examples in Section \ref{sec:topological}, the case of topological operators where the polynomial dependence on the external coordinates is just a constant (up to a sign-dependent factor).\par
		    We are then left with an analytic integral over the entire real $z$-line.
			\begin{align}
				I(x_i)&= \frac{2\pi i }{2} \int_{-\infty}^{+\infty} dz z^{n-2}  \sum_{j=1}^{n-1}	\frac{1}{2 i z\Pi_{i\neq j}(x_{ij}^2+2izx_{ij})}\\
				&=\frac{\pi  }{2} \int_{-\infty}^{+\infty} dz z^{n-3}  \sum_{j=1}^{n-1}	\frac{1}{\Pi_{i\neq j}(x_{ij}^2+2izx_{ij})}.
			\end{align}
			The integrand has a good large-$|z|$ behaviour and we can therefore analytically continue $z$ and evaluate this integral by contour integration, neglecting the vanishing contribution from the contour at $\infty$. There are poles at positions 
			\begin{align}
				z^*=-\frac{x_{ij}}{2i},
			\end{align} 
			with residues
			\begin{align}
				\hspace{-.2cm} \! Res_{z=z^*}\left(\frac{z^{n-3} }{\Pi_{k\neq j}(x_{kj}^2+2izx_{kj})}\right)
				=	\frac{x_{ij}^{n-4} }{(2i)^{n-4}\Pi_{k\neq j\neq i}x_{ik}x_{jk}}.
			\end{align}
			We close the contour in the UHP, in which only the poles where $x_i-x_j>0$ contribute.
			This gives the final result for odd-$n$ and ordered $x_i$
			\begin{align}\label{Eq:massless odd result}
				I_{n\,odd}(x_i)	&=\frac{\pi^2  }{ 2(2i)^{n-3}} \sum_{i>j}^{n}\frac{x_{ij}^{n-4}}{\Pi_{k\neq j\neq i}x_{ik}x_{jk}}.
			\end{align}
			This formula agrees with the canonical case of $n=3$, and the explicit results for $n=5$ and $n=7$ are given in Appendix \ref{App: list of correlators}.
			\subsubsection*{Even and odd $n$}
			For a generic number of external fields $n$, the integral
			\begin{align}
				I(x_i) &=\pi \int_0^\infty dz z^{n-3}\sum_{j=1}^{n}	\frac{1}{\Pi_{1=1,i\neq j}^{n}(x_{ij}^2+2izx_{ij})},
			\end{align}
			cannot be evaluated with contour integration. It can still be evaluated explicitly with the pole-matched, partial fraction decomposition of the integrand
			\begin{align}
				\hspace{-.4cm} \sum_j \frac{z^{n-3}}{\Pi_{k \neq j}\left(2ix_{kj}(z-i\frac{x_{kj}}{2})\right) } 	&=\frac{1}{(2i)^{n-2}} \sum_{i\neq j}\frac{x_{ij}^{n-4}}{\Pi_{k\neq i\neq j}x_{ik}x_{jk}}\frac{-1}{(z+a_{ij})},
			\end{align}
			where
			\begin{align}
				a_{ij} = \frac{x_{ij}}{2i}.
			\end{align}
			Using this decomposition, we obtain logarithm functions whose branch cut is chosen to be on the negative real axis. The choice of the branch of the logarithm is arbitrary since we do not cross any branch cut in the definite integration.\footnote{The author thanks Luke Corcoran for a discussion on this point.}
		    The convergent commuting of the sum and the integral is ensured by only taking the upper bound $\Lambda$ to infinity at the end of computations. This gives the result
			\begin{align}
				I(x_i)&=\lim_{\Lambda\rightarrow\infty}\frac{-\pi}{(2i)^{n-2}}\sum_{i\neq j}\frac{x_{ij}^{n-4}}{\Pi_{k\neq i\neq j}x_{ik}x_{jk}}\left(\log(a_{ij}+\Lambda)-\log(a_{ij})\right),
			\end{align}
			which can be simplified by averaging over the permutation of the two indices, since the sum is indiscriminate in $i$ and $j$. The first consequence is that the divergent term cancels in both cases, since we have
			\begin{align}
				\log(\Lambda) \sum_{i\neq j}\frac{x_{ij}^{n-4}}{\Pi_{k\neq i\neq j}x_{ik}x_{jk}}=0\qquad n\textrm{ even},
			\end{align}
			and in the odd-$n$ case we have a vanishing leading term since
			\begin{align}
				\log(\Lambda-i\frac{x_{ij}}{2})-\log(\Lambda+i\frac{x{ij}}{2})\xrightarrow{\Lambda\rightarrow\infty}0. 
			\end{align}
			Thus, we can write the result as
			\begin{align}\label{Eq: massless}
				I(x_i)&=\frac{\pi}{(2i)^{n-2}}\sum_{i\neq j}\frac{x_{ij}^{n-4}}{\Pi_{k\neq i\neq j}x_{ik}x_{kj}}\log\left(\frac{x_{ij}}{2i}\right),
			\end{align}
			which is a real quantity for  both the even case
			\begin{align}\label{Eq: result massless even}
				I_{even}(x_i)&=\frac{\pi}{2(2i)^{n-2}}\sum_{i\neq j}\frac{x_{ij}^{n-4}}{\Pi_{k\neq i\neq j}x_{ik}x_{jk}}\log\left(x_{ij}^2\right) ,
			\end{align}
			and the odd-$n$ case
			\begin{align}\label{Eq: result massless odd}
				I_{odd}(x_i)&=\frac{\pi}{2(2i)^{n-2}}\sum_{i\neq j}\frac{x_{ij}^{n-4}}{\Pi_{k\neq i\neq j}x_{ik}x_{jk}}\left(\log(a_{ij})-\log(-a_{ij})\right)\nonumber \\
				&=	\frac{\pi}{2(2i)^{n-2}}\left(i\pi \sum_{i>j}\frac{x_{ij}^{n-4}}{\Pi_{k\neq i\neq j}x_{ik}x_{jk}}-i \pi \sum_{i<j}\frac{x_{ij}^{n-4}}{\Pi_{k\neq i\neq j}x_{ik}x_{jk}}\right)\nonumber \\
				&=	\frac{\pi^2 }{2(2i)^{n-3}}\sum_{i>j}\frac{x_{ij}^{n-4}}{\Pi_{k\neq i\neq j}x_{ik}x_{jk}}.
			\end{align}
			Thus equation \eqref{Eq: massless} is consistent with the result \eqref{Eq:massless odd result} found in the previous section. The correlator \eqref{Eq: solution massless all n} follows from \eqref{Eq: result massless odd} and \eqref{Eq: result massless even}.
			This matches known literature for the case of the four-point functions, for example:
			\begin{align}\label{Eq: four-point}
				I_{\Delta=1,n=4} =-\frac{\pi }{2}\left(  \frac{\log \left(\chi_1\right)}{ 1-\chi_1}+\frac{\log \left(1-\chi_1\right)}{ \chi_1}\right)
			\end{align}
		and more cases are listed in Appendix \ref{App: list of correlators}.
			
			\subsection{Massive scalar fields }
			\label{subsec:massive }
			The method used in Section \ref{subsec:massless} is very powerful in the generic $n$ case but quickly increases in complexity when $\Delta>1$. However, another method can be used to obtain the massive $n$-point functions from the massless cases, as seen below in subsection \ref{Sec: pinching}.\par  For $\Delta=2$, the result can still be computed with this method relatively efficiently. The integral 
			\begin{align}\label{Eq: Integral Delta=2}
				I_{\Delta=2}(x_i) =\int dz z^{2n-2} \int dx\frac{ 1}{\Pi_{i=1}^{n}(z^2+(x-x_i)^2)^2}.
			\end{align}
			is evaluated by contour integration for the $x-$integral and partial fraction decomposition for the $z-$integral. Double poles lead to the less compact formula 
			\begin{align}\label{Eq Delta=2 result}
				&I_{\Delta=2,n}=\sum_i\sum_{j\neq i}\frac{-\pi}{2(2i)^{2n-4}x_{ij}^2}\partial_{x_j}\left( \frac{x_{ji}^{2n-5}}{\prod_{k\neq j,k\neq i }x_{ik}^2x_{jk}^2}\log\frac{x_{ji}}{2i}\right)\nonumber \\
				&+\sum_i \sum_{j\neq i}\partial_{x_i}\frac{-\pi}{(2i)^{2n-2}x_{ij}^2}\partial_{x_j}\left( \frac{x_{ji}^{2n-4}}{\prod_{k\neq j,k\neq i}x_{ik}^2x_{jk}^2} \log\frac{x_{ji}}{2i}\right),
			\end{align}
		which is derived in Appendix \ref{App: Delta=2 derivation}.
			One expects a similar structure at higher $\Delta$, where we have a double sum over the external coordinates $x_{i,j}$ and $\partial^{2\Delta}$ derivatives and $\Delta$ terms. Some evidence of this is the pinching presented in Section \ref{Sec: pinching} though subtleties in the order of limits prevent a general analysis in this paper. As such, the residue method loses its efficiency as we increase the dimension of the external operators.
			\subsection{Pinching}\label{Sec: pinching}
			One of the ways to relate correlators with differing number of points is through \textit{pinching}, that is, bringing an operator near another
			\begin{align}
				\lim_{x_i\rightarrow x_{i+1}}\langle\phi(x_1)...\phi(x_i) \phi(x_{i+1})...\phi(x_n) \rangle .
			\end{align}
			From the operator product expansion, one expects a divergence in this pinching limit. The contribution from the exchanged identity, in particular, leads to a power divergence
			\begin{align}
				\lim_{\epsilon\rightarrow0} \langle \phi_\Delta(x_1)\phi_\Delta(x_1+\epsilon)\rangle  \sim \epsilon^{-2\Delta}.
			\end{align}
		Useful results can still be obtained through a similar limit relating not the full correlators but the individual contact diagrams. In particular, the limit of the unit normalised propagators 
			\begin{align}
				\lim_{x_2\rightarrow x_1} \, \, \tilde{K}_{\Delta_1}(x_1;x,z)\tilde{K}_{\Delta_1}(x_1;x,z) = \tilde{K}_{\Delta_1+\Delta_2}(x_1;x,z)
			\end{align}
			indicates that the pinching of operators should relate higher-point integrals to higher-weight integrals if this limit commutes with the integral.
			\begin{figure}[h]
				\centering
				\begin{tikzpicture}[scale=.6]
					\def\x{4}
					\filldraw[blue!10!white] (0,0) circle (\x cm);
					\draw[] (0,0) circle (\x cm);

					\draw[] (0,0) to (\x, 0*\x);
					\node[anchor= west] at (\x, 0*\x){$x_5$};
					\draw[] (0,0) to (.5*\x, 0.866*\x);
					\node[anchor=south west] at (.5*\x, 0.866*\x){$x_6$};
					\draw[] (0,0) to (-.5*\x, 0.866*\x);
					\node[anchor=south east] at (-.5*\x, 0.866*\x){$x_1$};
					\draw[] (0,0) to (-1*\x, 0*\x);
					\node[anchor= east] at (-1*\x, 0*\x){$x_2$};
					\draw[] (0,0) to (-.5*\x, -0.866*\x);
					\node[anchor= north east] at (-.5*\x, -0.866*\x){$x_3$};
					\draw[] (0,0) to (.5*\x, -0.866*\x);
					\node[anchor=north west] at (.5*\x, -0.866*\x){$x_4$};
					
					\filldraw[]  (0,0) circle (0.7*\x pt);
				\end{tikzpicture} \hspace{1cm}
				\begin{tikzpicture}[scale=.6]
					\def\x{4}
					\filldraw[blue!10!white] (0,0) circle (\x cm);
					\draw[] (0,0) circle (\x cm);

					\draw[] (0,0) to (.5*\x, 0.866*\x);
					\node[anchor=south west] at (.5*\x, 0.866*\x){$x_{6}$};
					\draw[] (0,0) to (-.5*\x, 0.866*\x);
					\node[anchor=south east] at (-.5*\x, 0.866*\x){$x_1$};
					\draw[very thick] (0,0) to (-1*\x, 0*\x);
					\node[anchor=east] at (-1*\x, 0*\x){$x_2$};
					\draw[very thick] (0,0) to (\x, 0*\x);
					\node[anchor= west] at (\x, 0*\x){$x_5$};
					
					\filldraw[]  (0,0) circle (0.7*\x pt);
				\end{tikzpicture} 
				\caption{Pinching of the $I_{\Delta=1, n=6}$ integral to $I_{[1,2,2,1]}$ where the thick lines represent $\Delta=2$ bulk-to-boundary propagators.}
				\label{Fig: pinching}
			\end{figure}
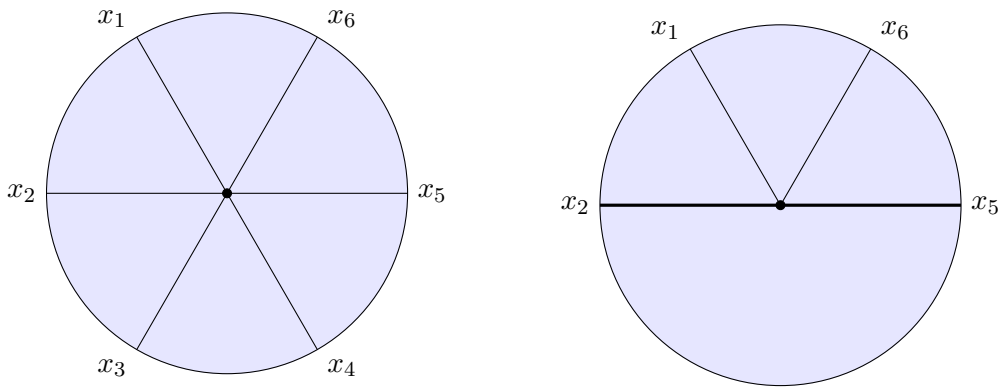\par
			One expects this commuting between the limit and the integral to break down as soon as one encounters divergences. In other words, in the absence of divergences, 
			\begin{align}
				\lim_{x_{i+1}\rightarrow x_{i}} I_{[\Delta_1,...,\Delta_n]}(x_1,...,x_n) = I_{[\Delta_1,...,\Delta_i+\Delta_{i+1},\bar{\Delta}_{i+1},...,\Delta_n]}(x_1,...,\bar{x}_{i+1},...,x_n),
			\end{align}
			where $\bar{x}_{i+1}$ denotes the absence of an operator at position $x_{i+1}$. The process can be iterated to form massive contact $n$-point diagrams from the basis of massless contact diagrams. 
			This can, in principle, be done for all values of $\Delta$ and $n$. A list of examples for scalars of differing dimensions is given in Appendix \ref{App: list of correlators}, agreeing with numerical integration and known results. This provides a non-trivial check of the 6-point function as well as a way to evaluate the four-point correlator of massive fields. 
			When divergences are present in the individual diagrams, these might cancel in the full correlator, and if not need to be regularised. \par 
			The simplest example in which divergences naturally appear in a pinched contact diagram is when considering the pinching from a four- to a three-point function
			\begin{align}
				\lim_{x_2\rightarrow x_1} <\phi(x_1)\phi(x_2)\phi(x_3)\phi(x_4)> &= 	\lim_{x_2\rightarrow x_1}\left( \frac{1}{x_{13}^2x_{24}^2} \left( \frac{-2\log(\chi)}{1-\chi}-\frac{2\log(1-\chi)}{\chi}\right)\right)\nonumber\\
				&=\frac{1}{x_{13}^2x_{14}^2}\lim_{\epsilon\rightarrow 0}\left(-2\log(\epsilon)-2\right),
			\end{align}
			where the pinched cross-ratio $\chi \rightarrow 0$ generates a divergence in the pinched correlator. 
			In some physical systems, the cancellation of such divergences can occur thanks to the symmetries of the theory. For example, in the pinching of $\mathcal{N}=4$ fields in \cite{Barrat:2021tpn}, the contraction of the R-symmetry indices with a null vector ensures that the protected operators form a chiral ring, and the powers of a single protected operator are still protected. In the generic case where the divergences are retained, these do not necessarily match the corresponding correlator.

			From the examples shown in this paper, it seems that the class of scalar contact diagrams follows this general property. In particular, since the dictionary of $D$-functions is well known, this provides a non-trivial check of the higher-point functions. One would be tempted to apply this pinching to the formal expression \eqref{Eq: massless} to have an independent derivation of the $\Delta=2$ case in \eqref{Eq Delta=2 result}. However, the pinching has to be done after the sum to have consistent limits, which impedes deriving the $\Delta=2$ case from the massless one. With this caveat in mind, the $n$-point contact diagrams of massless scalars can generate all $n$-point contact diagrams. \par
			An simplification which occurs in this setting is that one can bring the final point to $\infty$ before beginning the procedure detailed above.  In such a situation, the integral
			\begin{align}
				\lim_{x_n\rightarrow \infty}I_{\Delta_1,\Delta_2,...\Delta_n}(x_1,..,x_n) = \int dz dx z^{\Sigma-2}\prod_{i=1}^{n-1}\left(\frac{1}{z^2+(x-x_i)^2}\right)^{\Delta_i}
			\end{align}
			has a large $|z|$ behaviour as $\sim z^{2\Delta_n-\Sigma-2}$ and converges in the $z-$coordinate if
			\begin{align}
				\Sigma-2\Delta_n > -1
			\end{align}
		where $\Sigma = \sum_{i}\Delta_i$. In this situation, the coordinate $x_n$ can be ignored and the pinching can be evaluated in a straightforward manner by considering a variation of \ref{Eq: massless}\footnote{For other cases, we can easily map it to a convergent case through cyclic permutations of the external coordinates. }
		\begin{align}
			I_{\Delta_1,...,\Delta_n}(\chi_i)&=\lim_{\epsilon\rightarrow0}\text{Pinch}\frac{\pi}{(2i)^{n-2}}\sum_{i\neq j}\frac{x_{ij}^{\Sigma-4}}{\Pi_{k\neq i\neq j}x_{ik}x_{kj}}\log\left(\frac{x_{ij}}{2i}\right),\\
				\text{Pinch} &= x_{k+\sum_{p=0}^l\Delta_p}\rightarrow \chi_l+(k-1) \epsilon
		\end{align}
		where the pinching limit is in point-splitting formalism, $x_i=\{x_1,...,x_{\Sigma-\Delta_n}\}$, $k$ ranges from $1$ to $\Delta_{l+1}$ and $l$ ranges from $0$ to $n-2$. To condense the notation, we used the conventions
		\begin{align}
			\chi_0&=0&\chi_{n-2}&=1&\Delta_0&=0
		\end{align}
	A mathematica notebook is attached to this submission which evaluates this pinching along with giving the appropriate conformal and numerical prefactor. This is relatively efficient at computing $n$-point contact diagrams of arbitrary weights. 
	\section{An application: topological correlators}
		\label{sec: examples}
			This example considers non-Abelian gauge theories in AdS$_2$ and is an alternative construction to the Witten diagram computation in Appendix A.2 of \cite{Mezei:2017kmw}. For consistency with the notation in \cite{Mezei:2017kmw}, we denote the boundary coordinate by $t$ instead of $x$.
		\subsection*{Yang-Mills in AdS${}_2$ }
			\label{sec:topological}

			We review the setting of \cite{Mezei:2017kmw} where the strong coupling action is that of Yang-Mills theory in AdS$_2$ completed with a regulating boundary term
			\begin{align}
				S_{YM} &= \frac{1}{2g_{YM}^2}\int_{AdS_2} dx^2 \sqrt{-g}\textrm{Tr}\left(F_{\mu \nu}F^{\mu \nu}\right)\\
				S_{b^{y}}&=\frac{1}{g_{YM}^2}\int_{\partial AdS_2} dx \sqrt{-\gamma}\textrm{Tr}\left( A_i A^i-2A^iF_{\mu i}n^\mu \right),
			\end{align}
			where $\mu, \nu$ are the indices in the bulk coordinates of AdS$_2$, $i$ those of the boundary coordinates, and $n^\mu$ is a unit vector normal to the boundary of AdS$_2$. \par
\begin{figure}[H]
				\center
				\begin{tikzpicture}[scale=.8]
					\draw[->] (0,-2) to (0,5);
					\draw[->] (-7,0) to (7,0);
					\draw [thick, red, ->] (-6,.3) to (2,.3);
					\draw [thick, red] (2,.3) to (6,.3);
					\filldraw[red] (1,1) circle (2pt);
					\node[anchor= south ] at (2,1)  {$i (t_2-t_1)$};
					\filldraw[red] (1,-1) circle (2pt);
					\node[anchor= north ] at (2,-1)  {$-i(t_2-t_1)$};
					\draw [,thick,red] (-6,.3) to[out=90,in=180] (0,4) to[out=0,in=90](6,.3) ;
					\draw[] (6,5)to(6,4.5) to(7,4.5);
					\node[anchor = south west] at (6,4.5){$z\in \mathbb{C}$};
					\node[anchor= south] at (5,.3) {$\mathcal{C}_{\mathbb{R}}$};
					\node[anchor= south] at (3.5,3.5) {$\mathcal{C}_{\infty}$};
				\end{tikzpicture}
				
				\caption{Contour used for the $z-$integral in equation \eqref{Eq: z-integral topo}, the origin of the topological factor sgn$(t_1-t_2)$ is clear in this setup. The contour is closed in the UHP (the same analysis holds for the LHP closing). The pole contained within this contour depends on the sign of $(t_1-t_2)$ where, in this example, we have shown the case $t_1<t_2$.}
				\label{Fig: contour integral 3}
			\end{figure}
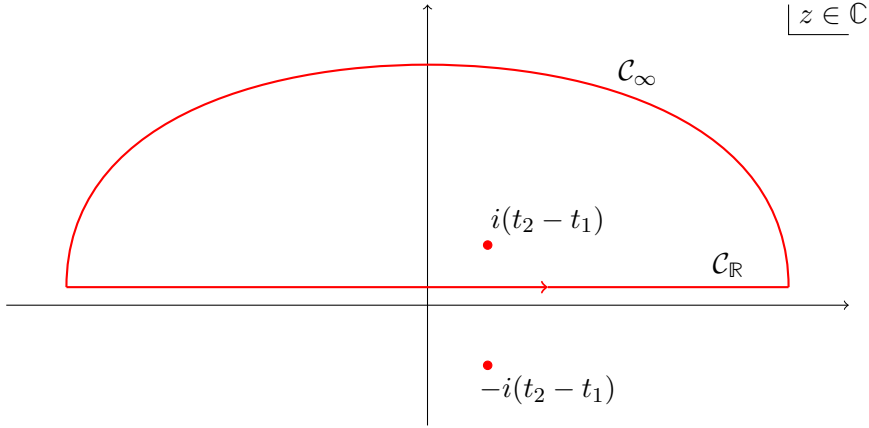

 
			In radial coordinates, the equation of motion is solved by
			\begin{align}
				F_{r \varphi} &= Q\sinh r & A_{\varphi} &= Q(\cosh r -1)&A_r&=0,
			\end{align}
			where $Q = Q_aT^a$ is an element of the Lie algebra of the theory and in the following, indices $a,b,a_i$ are those of the gauge algebra. This gives the on-shell action
			\begin{align}\label{Eq:on-shell action}
				\left(S_{tot}\right)_{\textrm{on-shell}} = -2\pi \frac{\textrm{Tr}(Q^2)}{g_{YM}^2}.
			\end{align}
			To relate the boundary fields to the bulk fields, the variation of the bulk action needs to be written in terms of the variation of the boundary field
			\begin{align}
				\delta S_{tot} &= \frac{2}{g_{YM}^2} \int_{\partial B}dx \sqrt{-\gamma}\textrm{Tr}\left( A^i \delta a\right)	& a &= \lim_{x^\mu\rightarrow \partial B}\left( A_i -F_{\mu i}n^\mu\right),
			\end{align}
			where $a$ is thus the corresponding boundary field and $i$ is the index corresponding to the boundary coordinate ($t$ in the following).
			The on-shell action \eqref{Eq:on-shell action} can be written in terms of the boundary fields $a$ through the equation
			\begin{align}
				a(\varphi) &= -u Q u^{-1}+iu \partial_\varphi u^{-1}\\
				u_0Qu_0^{-1} &= \frac{i}{2\pi}\log \left( P\exp\left(i\int_0^{2\pi}d\varphi a(\varphi)\right)\right)\label{Eq : uQ u},
			\end{align}
			where the $\varphi-$dependant large gauge transformations at the boundary are parametrized by $u$ and $P \exp$ denotes the usual path ordered exponential. The expression for the on-shell action is then proportional to the trace of \eqref{Eq : uQ u} squared,
			\begin{align}
				\Tr(Q^2) = 	\Tr((u_0 Q u_0^{-1})^2)  = -\frac{1}{4\pi^2}\Omega(a).
			\end{align}
			The expression for $\Omega(a)$ is a standard result in quantum mechanics and is solved by the Magnus expansion \cite{Wilcox:1967zz,2009PhR...470..151B}
			\begin{align}
				\exp\left(\Omega\right)= P \exp\left(i\int d\varphi a(\varphi))\right).
			\end{align}
			This can be used to find the dual correlators through the holographic dictionary
			\begin{align}
				<j^{a}(\varphi_1)j^{b}(\varphi_2)> &= \frac{\delta^{ab}}{4\pi g_{YM}^2}\\
				<j^{a}(\varphi_1)j^{b}(\varphi_2)j^{c}(\varphi_3)>& =-\frac{f^{abc}\textrm{sgn}{\varphi_{12}\varphi_{23}\varphi_{31}}}{4\pi g^2_{YM}}\\
				<j^{a_1}(\varphi_1)j^{a_2}(\varphi_2)j^{a_3}(\varphi_3)j^{a_4}(\varphi_4)>& =-\frac{f^{a a_1 a_2}f^{a a_3 a_4}}{4\pi g^2_{YM}}\left(\textrm{sgn}{\varphi_{12}\varphi_{24}\varphi_{43}\varphi_{31}}-\textrm{sgn}{\varphi_{21}\varphi_{14}\varphi_{43}\varphi_{32}}\right)\nonumber \\
				&\quad +(2\leftrightarrow 3)+(2\leftrightarrow 4).
			\end{align}
			where the indices $a,b,c, a_i$ are those of the gauge algebra. Through Witten diagrams, these correlators of boundary terms can be computed explicitly using the contour integral method detailed above. 
			The bulk-to-boundary propagators in Poincaré coordinates for the gauge field $A_\mu$ are \cite{DHoker:1999bve, Costa:2014kfa} 
			\begin{align}
				G_{\mu}(z,t;t_i) &= \frac{z^2+(t-t_i)^2}{2\pi z}\partial_\mu \left(\frac{t-ti}{z^2+(t-t_i)^2}\right),
			\end{align}
			or explicitly
			\begin{align}
				G_{z}(z,t;t_i) &=\frac{t_i-t}{\pi  \left((t-t_i)^2+z^2\right)}&G_t(z,t,t_i)&=\frac{z^2-(t-t_i)^2}{2\pi z(t-t_i)^2+z^2}.
			\end{align}
			The on-shell action is a pure boundary term
			\begin{align}
				S_{on-shell} &= \frac{1}{2g_{YM}^2}\int_{AdS_2} dx^2\sqrt{-g}\textrm{Tr}\left(D_\mu A_\nu F^{\mu \nu}\right)+\frac{1}{g_{YM}^2}\int _{\partial AdS_2}dx\sqrt{-\gamma} \textrm{Tr}\left(A_i A^i-2A^i F_{\mu i}n^\mu\right)\nonumber \\
				&=\frac{1}{g_{YM}^2}\int _{\partial AdS_2}dx\sqrt{-\gamma} \textrm{Tr}\left(A_i A^i+A^i F_{i \mu }n^\mu\right).
			\end{align}
			Explicitly, in the $(z,t)$ Poincaré coordinates, this gives\footnote{Note that the vector pointing out of the boundary goes in the $-z$ direction.}
			\begin{align}
				S_{on-shell} &=-\frac{1}{g_{YM}^2}\int dt z \textrm{Tr}\left(A_t A_t-z A_tF_{t z }\right)|_{z=0}.
			\end{align} 
			The two-point correlators are given by Wick contractions acting on this term
			\begin{align}
				<a^a(t_1)a^b(t_2)>&=\lim_{z\rightarrow 0}- \frac{1}{2g_{YM}^2}\int dt \delta^{ab} zG_t(z,t,t_2)\left(G_t(z,t;t_1)+z\partial_{[z}G_{t]}(z,t;t_1)\right) \\
				&=\lim_{z\rightarrow 0} \frac{(t_1-t_2)^2\delta^{ab}}{4 \pi g_{YM}^2 \left((t_1-t_2)^2+4 z^2\right)}\\
				&= \frac{\delta^{ab}}{4\pi g^2_{YM}}.
			\end{align}
			The three-point vertex is 
			\begin{align}
				S_3 =- \frac{1}{g_{YM}^2}\int dt dz z^2 f_{abc}A^a_zA^b_t\partial_{[z}A^c_{t]},
			\end{align}
			which gives a correlator
			\begin{align}
				\langle a^{a}(t_1)a^{b}(t_2)a^{c}(t_3)\rangle  = \frac{1}{ g_{YM}^2}\text{Perm}\left(f^{abc}I(t_1,t_2,t_3)\right),
			\end{align}
			where we define the single-Wick-contracted integral
			\begin{align}
				I(t_1,t_2,t_3)	 = \int dt dz z^2 G_z(z,t;t_1) G_t(z,t;t_2)\partial_{[z}G_{t]}(z,t;t_3).
			\end{align}
			The anti-symmetrised derivative removes the $t_3$ dependence, and the parity of this integrand under $z\rightarrow -z$ is the same as that of the odd $n$ massless scalar case (see section \ref{subsec:massless}) so we can evaluate both the $z$ and $t$ integrals with a complex contour.\footnote{There is a convergence issue in $I({t_1,t_2,t_3})$ which is solved when considering the sum of the Wick contractions, since the leading term is $t^{-1}$, this is always cancelled by an odd permutation of the indices $(1,2,3)$, the next to leading term is convergent.}
			Extending the $z$ variable to the entire real line we have
			\begin{align}
				I({t_1,t_2,t_3})&= \int_{-\infty}^{\infty}dt \int_0^{\infty} dz\frac{(t-t_1) \left((t-t_2)^2-z^2\right)}{4\pi ^3 z \left((t-t_1)^2+z^2\right) \left((t-t_2)^2+z^2\right)}\\
				&=\frac{1}{2} \int_{-\infty}^{\infty}dt \int_{-\infty}^{\infty} dz \, \, \textrm{sgn}(z)\frac{(t-t_1) \left((t-t_2)^2-z^2\right)}{4\pi ^3 z \left((t-t_1)^2+z^2\right) \left((t-t_2)^2+z^2\right)}.
			\end{align}
			This integral has a $t_i$-independent  contribution from the behaviour at $t\rightarrow \infty$. However, this is cancelled by the permutation and the antisymmetry of the structure constants. We will therefore ignore this contribution and evaluate the integral by contour integration. The $t-$integral evaluates to
			\begin{align}
				I({t_1,t_2,t_3}) &=\int_{-\infty}^\infty\frac{dz}{z}\frac{2 z t_{12}+i t_{12}^2+4 i z^2}{8 \pi ^2  \left(t_{12}^2+4 z^2\right)} \textrm{sgn}(z)^2\\
				&=\int_{-\infty}^\infty\frac{dz}{z}\frac{2 z t_{12}+i t_{12}^2+4 i z^2}{8 \pi ^2  \left(t_{12}^2+4 z^2\right)}  \label{Eq: z-integral topo}.
			\end{align}
			Just as in the previous case the factors of $\textrm{sgn}(z)$ cancel and leave an analytic function in $z$. This integral also has a pole at 0 and at $\infty$, these can also be cancelled using the antisymmetry of the structure constants of the algebra by considering a combination of Wick contractions; for example $	f^{a_1,a_2,a_3}\left(I_{t_1,t_2,t_3} -	I_{t_2,t_1,t_3}\right)  $. With this in mind, the integral can be evaluated using contour integration. The only remaining pole is at $z = \pm i \frac{t_1-t_2}{2}$ and therefore the integral will have a factor of $\textrm{sgn}(t_1-t_2)$ multiplying the residue at that point (see Figure \ref{Fig: contour integral 3}).\par 
This gives the result for one diagram
			\begin{align}
				I(t_1,t_2,t_3) = \frac{1}{2\pi}\textrm{sgn}(t_{12}),
			\end{align}
		and therefore
			\begin{align}
				\langle a^{a}(t_1)a^{b}(t_2)a^{c}(t_3)\rangle  = \frac{1}{8\pi g_{YM}^2}\left(\sum_{\sigma(\{1,2,3\})}\left(f^{a_1a_2a_3}\textrm{sgn}t_{12}\right)\right)|_{\{a_1,a_2,a_3\}\rightarrow \{a,b,c\}}.
			\end{align}
			Using the total antisymmetry of the structure constants we have 
			\begin{align}
				\sum_{\sigma(\{1,2,3\})}\left(f^{a_1a_2a_3}\textrm{sgn}(t_1-t_2)\right)|_{\{a_1,a_2,a_3\}\rightarrow \{a,b,c\}} = -2f^{abc}\, \textrm{sgn}(t_{12}t_{23}t_{31}),
			\end{align}
			which gives the final result
			\begin{align}
				\langle a^{a}(t_1)a^{b}(t_1)a^{c}(t_1)\rangle  = -\frac{1}{4\pi g_{YM}^2} f^{abc}\, \textrm{sgn}(t_{12}t_{23}t_{31}).
			\end{align}
			This agrees with the result for the topological three-point function seen above through a simple change of coordinates.\footnote{For higher-point functions, there is the subtlety that the boundary field $a$ is not the boundary limit of the gauge field $A_\mu$, but rather has a dependence on both $A_\mu$ and $F_{\mu \nu}$. This implies that the bulk-to-boundary propagator receives corrections from multi-source terms. These questions are addressed in \cite{Mezei:2017kmw} but are beyond the scope of this paper.}

		\section{Exchanges and Polyakov blocks}
		\label{sec:polyakov}
			
			Exchange diagrams immediately increase the difficulty of the calculation by adding a bulk integration as well as a bulk-to-bulk propagator. For example, the $t$-channel exchange diagram is given by the integral
			\begin{align}\label{Eq: Exchange integral}
				J(x_1,...,x_4) &= \int \frac{d^2z_a}{z_{a,0}^2}I(z_a,x_2,x_3)K_{\Delta_1}(x_1,z_a)K_{\Delta_4}(x_4,z_a)\\
				I(w,x_2,x_3) &= \int \frac{d^2z}{z_0^2}G_{\Delta_E}(w,z)K_{\Delta_2}(x_2,z)K_{\Delta_3}(x_3,z),
			\end{align}
			where $K_\Delta$ and $G_{\Delta_E}$ are defined in equations \eqref{Eq: bulk-to-boundary prop} and $z_a= (z_{a,0},z_{a,1})$ are bulk coordinates. 
			Using the isometries of AdS space, the exchange diagrams can be related to contact diagrams \cite{DHoker:1999mqo,Rastelli:2017udc} such as those presented in Section \ref{subsec:massless}. This toy example is slightly different to the case in \cite{DHoker:1999mqo}, since the sum of contact diagrams does not truncate, but allows for an explicit example of a non-vanishing Polyakov block. Since both bulk integrations in the exchange Witten diagram integral have conformal symmetry, the solution is invariant under the action of symmetry generators on the legs attached to each of the bulk points
			\begin{align}
				(\mathbf{\mathcal{L}}_a+\vec{L}_2+\vec{L}_3)I(z_a,x_a;x_2,x_3) = 0.
			\end{align}
			This allows one to relate the quadratic Casimir acting on the external legs to the Laplacian acting on the corresponding bulk point
			\begin{align}\label{Eq: casimir exchange}
				(C_{(23)}^{(2)}-m^2_E)I(z_a,x_a;x_2,x_3) &=  \int \frac{d^2z}{z_0^2}\left((\nabla_{a}^2-m_E^2)G_{\Delta_E}(w,z)\right)K_{\Delta_2}(x_2,z)K_{\Delta_3}(x_3,z).
			\end{align}
		 Given that the bulk-to-bulk propagator satisfies the equation of motion, this term in \eqref{Eq: casimir exchange} reduces to a delta function, thus reducing the number of integrals.
		The problem is then more tractable since the double AdS$_2$ bulk integrations are replaced by a differential equation relating the answer to the known case of contact diagrams,\footnote{A more detailed review of the derivation is presented in Appendix \ref{App: Polyakov}.} which is a single bulk integral. \par Let us now consider a $\lambda \phi^3$ interaction in AdS$_2$ which gives a non-vanishing three-point function and an exchange diagram for the four-point function. For the case of four-point correlators, the exchange diagram integral is solved explicitly in the $s$, $t$ and $u$ channels. 
			\begin{align}
				<\phi(x_1)\phi(x_2)\phi(x_3)\phi(x_4)> &=\frac{\lambda^2C_\Delta^4C_{\Delta_E}}{(x_{13}x_{24})^2}\left(f_t^{\Delta,\Delta_E}(\chi) +f_s^{\Delta,\Delta_E}(\chi) +f_u^{\Delta,\Delta_E}(\chi) \right),
			\end{align}
			where 
			$C_\Delta$ is the normalisation defined in equation \eqref{Eq: normalisation C_Delta}. For example in the t channel, 
\begin{align}\label{Eq: Casimir eq 1}
		(C_{(23)}^{(2)}-\Delta_E(\Delta_E-1))	 \frac{1}{(x_{13}x_{24})^{2\Delta_\phi}}  f_t^{\Delta,\Delta_E}(\chi)&= \frac{1}{(x_{13}x_{24})^{2\Delta_\phi}} I_{\Delta,n=4}(\chi),
\end{align}
			where $C_{(23)}^{(2)}$ is the quadratic Casimir acting on the external points 2 and 3, $\Delta_E$ is the conformal dimension of the exchanged operator, $f_t^{\Delta,\Delta_E}$ is the function of the cross-ratio $\chi$ corresponding to the Witten exchange diagram in the $t$-channel ($J(x_1,...,x_4)$ in equation \ref{Eq: Exchange integral}), and $I_{\Delta,n=4}$ is the contact integral defined in \eqref{Eq: contact integral}. For more details in the derivation and the computation, see Appendix \ref{App: Polyakov}. In one dimension, this differential equation simplifies to one of a single variable
			\begin{align}
			\hspace{-.3cm}	\left((\chi-1)\! \left((\chi-1) \chi f''(\chi)+(4 \Delta  \chi+\chi-1) f'(\chi)\right)\!+\! f(\chi) \left(2 \Delta  (2 \Delta  \chi-1)-m_E\right)\right)&=   I_{\Delta,n=4}(\chi).
			\end{align}
			And can be solved for example for $\Delta_\phi=\Delta_{exch}=1$
			\begin{align}
				f_{t}^{(1,1)} (\chi)&=\frac{\pi}{4}\frac{c_1+c_2 \log (\chi^2)+6 \text{Li}_3(\chi)- \text{Li}_2(\chi) \log (\chi^2)}{(\chi-1)^2}.
			\end{align}
			The same can be done in the other channels; in the $s$-channel, we have 
			\begin{align}
				f_{s}^{(1,1)}(\chi)&=\frac{\pi}{4}\frac{c_1+c_2\log \left((1-\chi)^2\right)+6 \text{Li}_3(1-\chi)- \text{Li}_2(1-\chi) \log\left( (1-\chi)^2\right)}{\chi^2}\label{Eq: fs1}\\
				&= f_t^{(1,1)}(1-\chi).
			\end{align}
			In the $u$-channel, we have 
			\begin{align}
				f_u^{(1,1)}(\chi)&= \frac{\pi}{4}\left(c_3+6 \text{Li}_3\left(\frac{\chi}{\chi-1}\right)- \text{Li}_2\left(\frac{\chi}{\chi-1}\right) \log \left((\frac{\chi}{1-\chi})^2\right)+c_4 \log \left((\frac{\chi}{1-\chi})^2\right)\right)\nonumber \\
				&= (1-\chi)^{-2}f_t(\frac{\chi}{\chi-1})
			\end{align} 
		    The symmetry of the 3 channels is clear:
			The $s$ and $t$ channels are related by $\chi\rightarrow 1-\chi$ crossing which equates their integration constants. The solution which is crossing-symmetric and makes the OPE expansion consistent\footnote{The full analysis requires the three-point diagram and is done in Appendix \ref{App: Polyakov}} has the integration coefficients equal to
			\begin{align}
				c_1=c_3 &= 6\zeta(3)\\
				c_2 = c_4 &= -\frac{\pi^2}{6} .
			\end{align}
			Additionally, this solution has the mildest Regge growth.\\
			We can then define the correlator from the sum of the exchanges in the different channels. 
			\begin{align}
				<\phi(x_1)\phi(x_2)\phi(x_3)\phi(x_4)> &=\frac{\lambda^2}{\pi^5(x_{13}x_{24})^2}\left(f_t^{\Delta,\Delta_E}(\chi) +f_s^{\Delta,\Delta_E}(\chi) +f_u^{\Delta,\Delta_E}(\chi) \right),
			\end{align}
			The sum of exchanged Witten diagrams can be related to the Polyakov block \cite{Mazac:2018qmi}, for an exchanged weight $\Delta=1$ and external weights $\Delta_\phi =1$,
			\begin{align}\label{Eq: First polyakov block}
				P_{1,1}^{(0)} (\chi)&= \frac{4}{\pi}\left(f_u^{(1,1)}(\chi) +f_t^{(1,1)}(\chi) +f_s^{(1,1)}(\chi) \right).
			\end{align}\par
			Notice that the $u$ and $t$ channels evaluated in the $\frac{\chi}{1-\chi}$ variable are well defined on the analytic continuation to the interval $0<\chi<1$. 
			\begin{align}
				f_u(\chi) = (1-\chi)^{-2}f_t(\frac{\chi}{\chi-1})\\
				f_t(\chi) = (1-\chi)^{-2}f_u(\frac{\chi}{\chi-1}).
			\end{align}
			Due to this `pseudo-braiding' and crossing properties of this analytically continued function, the double-discontinuity defined in \cite{Mazac:2018qmi} and reviewed in \eqref{Eq: Ddisc}, can be evaluated quite easily as
			\begin{align}
				dDisc^{(+)}[P_{(1,1)}(\chi)] = \frac{2 G_{h=1}(\chi)}{\chi^2}.
			\end{align}
			This is the discontinuity in the $s$-channel of the corresponding Polyakov block \cite{Mazac:2018qmi}. Additionally, the bosonic continuation defined via \eqref{Eq: bosonic continuation} is fully symmetric under $s\rightarrow t$ and $s\rightarrow u$, and is Regge-bounded. Therefore, $P^{(0)}_{1,1}$ in \eqref{Eq: First polyakov block} is the Polyakov block (defined on $0<\chi<1$)  with external weight $\Delta=1$ and exchanged weight $\Delta_E=1$,
			\begin{align}\label{Eq: Exchange solution Delta=1}
				P_{1,1}^{(0)} (\chi)&=\text{Li}_2\left(\frac{\chi}{\chi-1}\right) \log \left(\frac{\chi^2}{(\chi-1)^2}\right)-6 \text{Li}_3\left(\frac{\chi}{\chi-1}\right)-\frac{1}{6} \pi ^2 \log \left(\frac{\chi^2}{(\chi-1)^2}\right)+6 \zeta (3) \nonumber \\
				&+ \frac{\text{Li}_2(1-\chi) \log \left((\chi-1)^2\right)-6 \text{Li}_3(1-\chi)-\frac{1}{6} \pi ^2 \log \left((\chi-1)^2\right)+6 \zeta (3)}{\chi^2}\nonumber \\
				&+\frac{\text{Li}_2(\chi) \log \left(\chi^2\right)-6 \text{Li}_3(\chi)-\frac{1}{6} \pi ^2 \log \left(\chi^2\right)+6 \zeta (3)}{(\chi-1)^2} .
			\end{align}
			
			This agrees with the computation\footnote{We thank Pietro Ferrero for sharing results relating to \cite{Ferrero:2019luz} allowing for a verification of this result.} done via the conformal bootstrap in \cite{Ferrero:2019luz}. Similarly, higher exchanged weights or external weights can be computed, see Appendix \ref{App: Polyakov}. Along with constraints from the double-discontinuity and a suitable ansatz, this method might provide a way to compute all Polyakov blocks  $P_{1,\Delta_E}^{(0)} (\chi)$.

		\section*{Acknowledgements}
		The author specially thanks Lorenzo Bianchi, Valentina Forini and the referees for their comments and feedback and would also like to thank Julien Barrat, Davide Bonomi, Olivia Brett, Luke Corcoran, Pietro Ferrero,  Luca Griguolo, Luigi Guerini, Carlo Meneghelli and Giulia Peveri for useful discussions. The author also warmly thanks the Dipartimento SMFI, Università di Parma where some of the research was conducted. The research received funding from the European Union’s Horizon 2020 research and innovation programme under the Marie Sklodowska-Curie grant agreement No 813942 "Europlex" and from  the Deutsche Forschungsgemeinschaft (DFG, German Research Foundation) - Projektnummer 417533893/GRK2575 ”Rethinking Quantum Field Theory”.
	
	
	\appendix
		
		\section{\texorpdfstring{$n$}{Lg}-point contact integrals}
		This appendix compiles the derivation of the $\Delta=2$ $n$-point contact integral, a non-exhaustive list of $n$-point $D$-functions, and a short numerical analysis of the results obtained for a large number of external legs.
		\subsection{Derivation of 	\texorpdfstring{$\Delta=2$}{Lg}}\label{App: Delta=2 derivation}
		One may of course be interested in higher $\Delta$. This is where this method loses some of its power. While it is very powerful in the generic $n$ regime, it quickly increases in complexity when $\Delta$ is increased. However, the complexity will only be combinatorial and not intrinsic.
		The integrand of
		\begin{align}
			I(x_i) =\int dz z^{2n-2} \int dx\frac{ 1}{\Pi_{i=1}^{N-1}(z^2+(x-x_i)^2)^2},
		\end{align}
		only has double poles (and single poles from the expansion around these poles), at the position $x = x_i+i z$ with residue
		\begin{align}
			&\underset{x = x_i+iz}{Res} \left(\! \frac{ 1}{\Pi(z^2+(x-x_i)^2)^2}\right) \!=\!\partial_x\left( \frac{1}{(x-(x_i-iz))^2}	\frac{ 1}{{\displaystyle \prod_{j,j\neq i}}(z^2+(x-x_j)^2)^2}\right)\Biggr|_{x=x_i+iz} \\
			&= -\sum_i	\frac{1}{4z^2}(\frac{1}{iz}+\partial_{x_i})\left( \frac{1}{\Pi_{j\neq i}x_{ij}^2(2i z +x_{ij})^2}\right),\nonumber
		\end{align}
		where the residue was massaged into a more usable form.
		As is the massless case, these can be integrated by using the partial fraction decomposition. By comparison of simple and double poles we have
		\begin{align}
			\sum_i \frac{z^{2n-n_0}}{\prod_{k}(z+a_{ik})^2} &= \sum_{i\neq j}\frac{c_{ij}}{(z+a_{ij})^2}-\sum_{i\neq j}\frac{\partial_{a_{ij}}c_{ij}}{(z+a_{ij})}\\
			&= -\sum_{i\neq j}\partial_{a_{ij}}\left(\frac{c_{ij}}{ (z+a_{ij})}\right),\\
			c_{ij}&= 	\frac{(-a_{ij})^{2n-n_0}}{\prod_{k\neq j}(a_{ki}-a_{ji})^2}. 
		\end{align}\bigskip
		as long as the $n_0>-1$. This is integrated by sight
		\begin{align}
			\int_0^\Lambda dz \left( \sum_{i\neq j}\frac{c_{ij}}{(z+a_{ij})^2}-\sum_{i}\frac{\partial_{a_{ij}}c_{ij}}{(z+a_{ij})}\right) =\sum_{i\neq j}\partial_{a_{ij}}\left(c_{ij}\log(a_{ij})\right)+(\partial_{a_i}c_{ij})\log(\Lambda).
		\end{align}
		\vspace{2mm} Explicitly, we have
		\begin{align}
			&\int  dz z^{2n-2}\frac{-1}{4z^2}\sum_i \partial_{x_i}\left( \frac{1}{\prod_{j\neq i}x_{ij}^2(2i z +x_{ij})^2}\right)\\
			&=\sum_i \sum_{j\neq i}\partial_{x_i}\frac{1}{4(2i)^{2n-3}x_{ij}^2}\partial_{x_j}\left( \frac{x_{ij}^{2n-4}}{\prod_{k\neq j,k\neq i}x_{ik}^2x_{jk}^2}\log\frac{x_{ji}}{2i}\right),
		\end{align}
		and 
		\begin{align}
			&\int dz \frac{z^{2n-5}}{4i}\left( \frac{1}{\prod_{j\neq i}x_{ij}^2(2i z +x_{ij})^2}\right)\nonumber\\
			&=\sum_i\sum_{j\neq i}\frac{1}{4i(2i)^{2n-4}x_{ij}^2}\partial_{x_j}\left( \frac{x_{ji}^{2n-5}}{\prod_{k\neq j,k\neq i }x_{ik}^2x_{jk}^2}\log\frac{x_{ji}}{2i}\right).
		\end{align}
		As in the massless case, the divergent terms cancel and the logarithm is a well-defined function with a branch cut on the negative real axis.
		This gives the result
		\begin{align}
			I_{\Delta=2,n}(x_i)&=\sum_i\sum_{j\neq i}\frac{-\pi}{2(2i)^{2n-4}x_{ij}^2}\partial_{x_j}\left( \frac{x_{ji}^{2n-5}}{\prod_{k\neq j,k\neq i }x_{ik}^2x_{jk}^2}\log\frac{x_{ji}}{2i}\right)\nonumber \\
			+\sum_i &\sum_{j\neq i}\partial_{x_i}\frac{-\pi}{(2i)^{2n-2}x_{ij}^2}\partial_{x_j}\left( \frac{x_{ij}^{2n-4}}{\prod_{k\neq j,k\neq i}x_{ik}^2x_{jk}^2}\log\frac{x_{ji}}{2i}\right).
		\end{align}

		\subsection{Library of contact correlators}\label{App: list of correlators}
		In the main body, results are naturally written in terms on the cross-ratios $u_i$ defined in equation \eqref{Eq: cross ratio}. However, they also hold for the  external coordinates, for example $\{x_1,...,x_4\}$ combine naturally to form the cross-ratio $u_1$ in \eqref{Eq: four-point} in the case of the four-point function 
		\begin{align}
			I_{\Delta=1,n=4}(x_1,..,x_4)&=-\frac{\pi}{2}\Big( \frac{\log x_{12}}{x_{23} x_{13}x_{24}x_{14}}+\frac{\log x_{13}}{x_{12}x_{23}x_{34}x_{14}}+\frac{\log x_{23}}{x_{12}x_{13}x_{34}x_{24}}\nonumber \\
			&+\frac{\log x_{34}}{x_{13}x_{23}x_{14}x_{24}}+\frac{\log x_{24}}{x_{12}x_{23}x_{14}x_{34}}+\frac{\log x_{14}}{x_{12}x_{13}x_{24}x_{34}}\nonumber \Big)\\
			&= -\frac{\pi}{2(x_{13}x_{24})^2}\left(\frac{x_{13}x_{24}}{x_{14}x_{23}}\log\left(\frac{x_{12}x_{34}}{x_{13}x_{24}} \right)  +\frac{x_{13}x_{24}}{x_{12}x_{34}}\log\left(\frac{x_{14}x_{23}}{x_{13}x_{24}} \right) \right).
		\end{align} 
		Below, we include a few examples $I(0,\chi_1,...,\chi_{n-3},1,\infty)$ of the contact integral \eqref{Eq: contact integral} evaluated in the cross-ratios defined in \eqref{Eq: cross ratio}, where we use the notation $I_{\Delta,n}$ for equal dimension operators and $I_{[\Delta_1,...,\Delta_n]}$ to include external operators of different dimensions.
		
		\begin{align}
			&I_{1,3}  = \frac{\pi ^2}{2}&&&&&\qquad \qquad&\\
			&I_{1,4} =-\frac{\pi }{2}\left(  \frac{\log \left(\chi_1\right)}{ 1-\chi_1}+\frac{\log \left(1-\chi_1\right)}{ \chi_1}\right)\\
			&I_{1,5} = \frac{\pi ^2}{4 \chi_2\left(1-\chi_1\right)}\\
			&I_{1,6}= \frac{\pi}{8}\left(\frac{\left(\chi_1-1\right){}^2 \log \left(1-\chi_1\right)}{\chi_1 \left(\chi_1-\chi_2\right) \left(\chi_2-1\right) \left(\chi_1-\chi_3\right) \left(\chi_3-1\right)}\right. \nonumber \\
			&\left.+\frac{\left(\chi_1-\chi_2\right){}^2 \log \left(\chi_2-\chi_1\right)}{\left(\chi_1-1\right) \chi_1 \left(\chi_2-1\right) \chi_2 \left(\chi_1-\chi_3\right) \left(\chi_2-\chi_3\right)}-\frac{\left(\chi_2-1\right){}^2 \log \left(1-\chi_2\right)}{\left(\chi_1-1\right) \left(\chi_1-\chi_2\right) \chi_2 \left(\chi_2-\chi_3\right) \left(\chi_3-1\right)}\right.\nonumber \\
			&\left. +\frac{\chi_2^2 \log \left(\chi_2\right)}{\chi_1 \left(\chi_1-\chi_2\right) \left(\chi_2-1\right) \left(\chi_2-\chi_3\right) \chi_3}+\frac{\left(\chi_3-1\right){}^2 \log \left(1-\chi_3\right)}{\left(\chi_1-1\right) \left(\chi_2-1\right) \left(\chi_1-\chi_3\right) \left(\chi_2-\chi_3\right) \chi_3}\right.\nonumber \\
			&\left.+\frac{\left(\chi_2-\chi_3\right){}^2 \log \left(\chi_3-\chi_2\right)}{\left(\chi_1-\chi_2\right) \left(\chi_2-1\right) \chi_2 \left(\chi_1-\chi_3\right) \left(\chi_3-1\right) \chi_3}-\frac{\left(\chi_1-\chi_3\right){}^2 \log \left(\chi_3-\chi_1\right)}{\left(\chi_1-1\right) \chi_1 \left(\chi_1-\chi_2\right) \left(\chi_2-\chi_3\right) \left(\chi_3-1\right) \chi_3}\right.\nonumber\\
			&\left.-\frac{\chi_3^2 \log \left(\chi_3\right)}{\chi_1 \chi_2 \left(\chi_1-\chi_3\right) \left(\chi_2-\chi_3\right) \left(\chi_3-1\right)}-\frac{\chi_1^2 \log \left(\chi_1\right)}{\left(\chi_1-1\right) \left(\chi_1-\chi_2\right) \chi_2 \left(\chi_1-\chi_3\right) \chi_3}\right)\\
			&I_{1,7} =\frac{\pi ^2}{16 \left(\chi_1-1\right) \left(\chi_2-1\right) \chi_2 \left(\chi_1-\chi_3\right) \left(\chi_3-1\right) \chi_3 \left(\chi_1-\chi_4\right) \left(\chi_2-\chi_4\right) \chi_4} \nonumber \\
			&\Big(\left(\chi_2-1\right) \left(\chi_3-1\right) \left(\chi_2-\chi_4\right) \chi_1^2\nonumber \\
			&+\chi_2 \left(\chi_3^2+\chi_4 \left(\chi_3+\chi_4-2\right) \chi_3-\chi_4-\chi_2 \left(\chi_3-1\right) \left(\chi_3+\chi_4+1\right)\right) \chi_1 \nonumber \\
			&+\chi_2 \left(\chi_2 \left(\chi_3-1\right) \left(\chi_3+\left(\chi_3+1\right) \chi_4\right)-\chi_3 \left(\chi_3 \left(\chi_4^2+\chi_4+1\right)-3 \chi_4\right)\right) \Big)
		\end{align}
		

		For the massive cases, we find agreement between the result of pinching and that of the formula \eqref{Eq Delta=2 result} which gives for the first few cases:
		\begin{align}
			&I_{2,3}= \frac{3 \pi }{8}\\
			&I_{2,4}=-\frac{\pi  ((\chi -1) \chi +1)}{8 (\chi -1)^2 \chi ^2}-\frac{\pi  \left(2 \chi ^2+\chi +2\right) \log (1-\chi )}{16 \chi ^3}+\frac{\pi  (\chi  (2 \chi -5)+5) \log (\chi )}{16 (\chi -1)^3}\\
			&I_{2,5} =\frac{\pi}{32} \big(\left. \right.\nonumber \\
			&+\frac{(\chi_1-\chi_2)^2 \left(\chi_1^3 (2 \chi_2-1)+\chi_1^2 (3 (\chi_2-2) \chi_2+2)+\chi_1 \chi_2 (2 (\chi_2-3) \chi_2+3)-(\chi_2-2) \chi_2^2\right) \log (\chi_2-\chi_1)}{(\chi_1-1)^3 \chi_1^3 (1-\chi_2)^3 \chi_2^3}\nonumber \\
			&+\frac{(\chi_2-1)^2 \left(\chi_1^2 (-(\chi_2 (2 \chi_2+3)+2))+\chi_1 (\chi_2+1) (\chi_2 (\chi_2+5)+1)-\chi_2 (\chi_2 (2 \chi_2+3)+2)\right) \log (1-\chi_2)}{(\chi_1-1)^3 \chi_2^3 (\chi_1-\chi_2)^3} \nonumber \\
			&+\frac{(\chi_1-1)^2 \left((\chi_1 (2 \chi_1+3)+2) \chi_2^2-(\chi_1+1) (\chi_1 (\chi_1+5)+1) \chi_2+\chi_1 (\chi_1 (2 \chi_1+3)+2)\right) \log (1-\chi_1)}{\chi_1^3 (\chi_2-1)^3 (\chi_1-\chi_2)^3}\nonumber \\
			&+\frac{\chi_2^2 \left(7 \chi_1^2-(\chi_1+1) \chi_2^3+(\chi_1+2) (2 \chi_1+1) \chi_2^2-7 \chi_1 (\chi_1+1) \chi_2\right) \log (\chi_2)}{\chi_1^3 (\chi_2-1)^3 (\chi_1-\chi_2)^3}\nonumber \\
			&+\frac{\chi_1^2 \left(\chi_1^3 (\chi_2+1)-\chi_1^2 (\chi_2+2) (2 \chi_2+1)+7 \chi_1 \chi_2 (\chi_2+1)-7 \chi_2^2\right) \log (\chi_1)}{(\chi_1-1)^3 \chi_2^3 (\chi_1-\chi_2)^3}\nonumber \\
			&+\frac{-2 \chi_1^4 ((\chi_2-1) \chi_2+1)+\chi_1^3 \left(2 \chi_2^3+\chi_2^2+\chi_2+2\right)-2 \chi_2^2 ((\chi_2-1) \chi_2+1)}{2 (\chi_1-1)^2 \chi_1^2 (\chi_2-1)^2 \chi_2^2 (\chi_1-\chi_2)^2} \nonumber \\
			&+\frac{\chi_1 \left(-2 \chi_2^4+\chi_2^3-6 \chi_2^2+\chi_2-2\right)+\chi_2 \left(2 \chi_2^3+\chi_2^2+\chi_2+2\right)}{2 (\chi_1-1)^2 \chi_1 (\chi_2-1)^2 \chi_2^2 (\chi_1-\chi_2)^2} \big)
		\end{align}
		From pinching, we define the integral with the pinched weights at positions $x_i$, e.g $I_{[1,2,2,1]}$
		\begin{align}
			I_{[2,2,2]}&= \frac{3 \pi }{8}\\
			I_{[1,1,1,2]}&= \frac{\pi ^2}{4 x_{13} x_{14} x_{24}^2 x_{34}}\\
			I_{[2,2,1,1]}&= -\frac{\pi  (\chi +2) \log (1-\chi )}{8 \chi ^3}+\frac{\pi }{8 \chi ^2(1-\chi)}+\frac{\pi  \log (\chi )}{8 (\chi -1)^2}\\
			I_{[2,1,2,1]}&=-\frac{(2 \pi  \chi +\pi ) \log (1-\chi )}{8 \chi ^2}+\frac{\pi }{8 (\chi -1) \chi }+\frac{\pi  (2 \chi -3) \log (\chi )}{8 (\chi -1)^2}\\
			I_{[1,2,2,1]}&=\frac{\pi  \log (1-\chi )}{8 \chi ^2}+\frac{\pi }{8 (\chi -1)^2 \chi }-\frac{\pi  (\chi -3) \log (\chi )}{8 (\chi -1)^3}\\
			I_{[2,2,2,1]}&= \frac{\pi ^2}{8 \chi (1-\chi)}\\
			I_{[3,1,2,1]}&= \frac{3 \pi ^2}{16 \chi }\\
			I_{[1,2,1,1,1]}&= -\frac{\pi  \chi_2^2 \log (\chi_2)}{8 \chi_1^2 (\chi_2-1) (\chi_1-\chi_2)^2}+\frac{\pi  (\chi_1 (\chi_1-2 \chi_2+2)-\chi_2) \log (1-\chi_1)}{8 \chi_1^2 (\chi_2-1) (\chi_1-\chi_2)^2}\nonumber \\
			&+\frac{\pi  (\chi_2-\chi_1 (\chi_1+2 \chi_2-2)) \log (\chi_2-\chi_1)}{8 (\chi_1-1)^2 \chi_1^2 (\chi_2-1) \chi_2}-\frac{\pi  \left(\chi_1^2-2 \chi_1 (\chi_2+1)+3 \chi_2\right) \log (\chi_1)}{8 (\chi_1-1)^2 \chi_2 (\chi_1-\chi_2)^2}\nonumber \\
			&+\frac{\pi  (\chi_2-1)^2 \log (1-\chi_2)}{8 (\chi_1-1)^2 \chi_2 (\chi_1-\chi_2)^2}\\
			I_{[2,1,1,2,1]}&= -\frac{\pi ^2 (2 \chi_1 \chi_2+\chi_1-3 \chi_2)}{16 (\chi_1-1)^2 \chi_2^2}\\
			I_{[1,1,1,1,2,1]}&= \frac{\pi ^2 \left(\chi_1^2 \left(\chi_2-1\right){}^2+\chi_2 \left(\chi_2+\left(2 \chi_2-3\right) \chi_3\right)+\chi_1 \chi_2 \left(2 \chi_3-\chi_2 \left(\chi_3+2\right)+1\right)\right)}{16 \left(\chi_1-1\right){}^2 \left(\chi_2-1\right){}^2 \chi_2 \left(\chi_1-\chi_3\right) \chi_3}
		\end{align}
	
	There are also some divergent cases
	\begin{align}
			I_{[2,1,1]}&= \frac{\pi}{2} (1- \log (\epsilon ))\\
			I_{[1,3,1,1]}&= \frac{5 \pi }{16 (\chi -1)^2 \chi ^2}-\frac{\pi  ((\chi -3) \chi +3) \log (\chi )}{8 (\chi -1)^3 \chi ^2}+\frac{\pi  \left(\chi ^2+\chi +1\right) \log (1-\chi )}{8 (\chi -1)^2 \chi ^3}\nonumber \\
			&\qquad-\frac{3 \pi  \log (\epsilon )}{8 (\chi -1)^2 \chi ^2}
	\end{align}
		We also find agreement between the pinching of the $2n$-point function of massless correlators and the $n$-point function of $\Delta=2$ correlators up to $n=8$, but these were omitted from the text since they are bulky and not elucidating.
		Notice that the prefactor \eqref{Eq: prefactor} has no neighbouring terms of  type $x_{i,i+1}$ except for the $x_{n-1,n}$ term and the $n=3$ cases, so no single pinching will lead to divergences in the prefactor. In general, one expects the divergences appearing in the pinching to be physical divergences which need to be considered and not artefacts of this method.
		\subsection{Numerical and analytical agreement}
\begin{figure}[h]
			\centering
			\includegraphics[scale=.8]{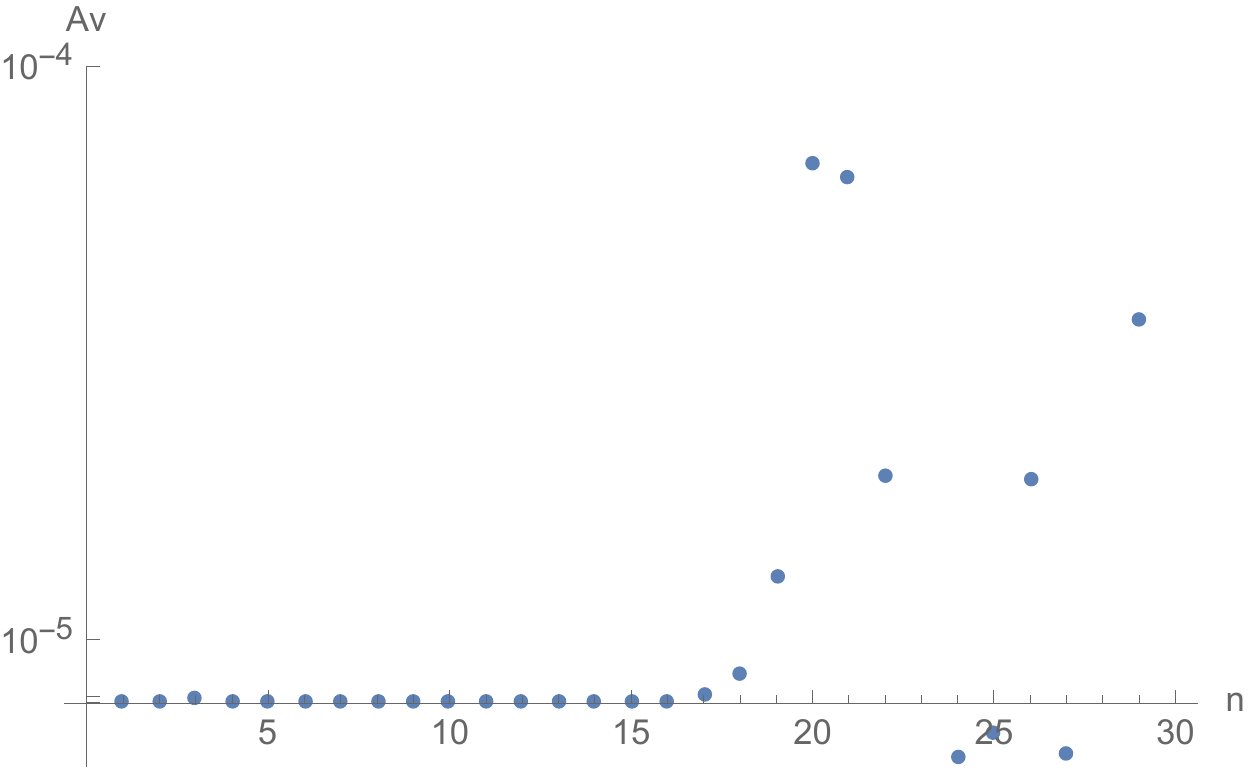}
			\caption{The quantity plotted (for clarity on a logarithmic scale) $\text{Av} = \langle\frac{I_n-I_{n,\text{num}}}{I_n}\rangle$ is the weighted difference between the numerical integration $I_{\Delta=1,n}$ in equation \eqref{Eq n-point numerics} and the corresponding analytical result, averaged over $0<\chi<1$ for a number of external points ranging from $n=3$ to $n=17$. For this range, the leading error comes from the machine precision in the integral near the endpoints and thus is of the same order for $n=3$ and $n=17$. For higher $n$, the numerical stability of the integral becomes problematic and the numerical results should not be trusted.}
			\label{fig:numerics-n7}
		\end{figure}

		An additional check for the validity of these results is to perform a numerical integration of these quantities for various values of $n$. This is done through fixing all but one parameter, for example considering the numerical integration of
		\begin{align}\label{Eq n-point numerics}
			I_{\Delta,n}(0,\chi,1,2,...,n-2) = \int dz dx\left( \frac{z}{(z^2+(x-\chi)^2)}\right)^{\Delta}\, \, \prod_{k=0}^{n-2}  \left( \frac{z}{(z^2+(x-k)^2)}\right)^{\Delta},
		\end{align}
		where $\chi$ is a free parameter which allows us to compare \eqref{Eq n-point numerics} to the analytic results in \eqref{Eq: massless}. The difference between the two can be seen in Figure \ref{fig:numerics-n7} and shows good agreement for $3<n<30$. In the figure, the normalised average over values in $0<\chi<1$ of the difference between \eqref{Eq n-point numerics} and the analytic result \eqref{Eq: massless} is plotted for varying numbers $n$ of external operators.
		
		A second check of validity can be made through analytical comparisons with known results. The results for $n=3$ and $n=4$ are well known but become sparser as $n$ grows larger, however, pinching also provides a way to compare higher$-n$ results to higher$-\Delta$ four-point correlators, this is done in Appendix \ref{App: list of correlators}.

		\section{Exchange diagrams}
		This Appendix contains complementary material relating to the derivation and interpretation of the exchange diagrams. First, details on the conformal Casimir and its relation to the equation of motion in AdS are given.  Then the derivation of the relation between exchange diagrams and contact diagrams from \cite{Rastelli:2017udc,Zhou:2020ptb} is reviewed. This is then applied to the toy model of $\phi^3$ interaction in AdS$_2$, which gives insight into Polyakov blocks whose perturbative strong coupling structure is shown explicitly
		\subsection*{Quadratic Casimir }\label{App: quad casimir}
		The $n$-Casimir of the boundary conformal group is given by
		\begin{align}
			C^{(n)}_{i_1..i_n} = \frac{1}{2}\{\sum_{k=1}^n L_{i_k}^{(0)},\sum_{k=1}^n L_{i_k}^{(0)}\}-\frac{1}{2} [\sum_{k=1}^n L_{i_k}^{-\alpha},\sum_{k=1}^n L_{i_k}^{\alpha}],
		\end{align}
		where $L^{(0)}$ are elements of the Cartan and the others are the simple roots. Explicitly for a $d=1$ conformal boundary the differential expression of the operators is:
		\begin{align}
			D &= L_0 = \Delta+x\partial_x& P &= L_{-1} = -\partial_x&	K& = L_{+1} = -2\Delta x-x^2\partial_x
		\end{align}
		This leads to a linear Casimir
		\begin{align}
			C^{(1)}_a &=  \Delta(\Delta-1),
		\end{align}
		which is the mass-squared of the bulk operator, and a quadratic Casimir:
		\begin{align}
			C^{(2)}_{x,y} &=2 (x-y) (- \Delta_1 \partial_y + \Delta_2 \partial_x )-(y-x)^2 \partial_x\partial_y+(\Delta_1+\Delta_2-1) (\Delta_1+\Delta_2).
		\end{align}
		The quadratic Casimir of the AdS$_2$ isometries is the Laplacian of AdS,
		this is best seen in flat embedding coordinates where the generators are given by 
		\begin{align}
			J_{AB} = -i(X_A\partial_B-X_B\partial_A),
		\end{align}
		where $X_A X^A=1$. The Quadratic casimir of the AdS isometries in embedding coordinates is then
		\begin{align}
			-\frac{1}{2}\mathcal{L}_a \mathcal{L}^a &=- \frac{1}{2}J_{AB}J^{AB}\\
			&= \frac{1}{2}(X_A\partial_B-X_B\partial_A)(X^A\partial^B-X^B\partial^A)\\
			&=X_AX^A\partial_B\partial^B +(1-d)X_B\partial^B\\
			&=\partial_A\partial^A,
		\end{align}
		which is the coordinate-independent Laplacian. 
		
		A case of interest in this paper is when we have $\Delta_1=\Delta_2$. The conformal quadratic Casimir then simplifies further to
		\begin{align}
			C^{(2)}_{z} = 2 \Delta  (2 \Delta -1) f(z)-z \left((z-1) z f''(z)+(2 \Delta  (z-2)+z) f'(z)\right),
		\end{align}
		where we have made a change of variable 
		\begin{align}
			z = 1-\frac{x}{y}. 
		\end{align}
		Such changes of variables can be made to reduce this differential equation into a single variable differential equation for each of the $(s,t,u)$ exchange channels.
		
		\subsection{Relating the exchange and contact diagrams}\label{Sec: relating exchange to contact diagrams}
		We review the analysis from \cite{1999NuPhB.562..395D,Zhou:2018sfz}, which goes through a detailed computation of the exchange diagram and the $z$ integral. However, they specialise in the case where the resulting exchange correlator can be written in terms of a finite sum of $D$-functions. The toy model considered in this paper does not satisfy the conditions needed for such a simplification, but still is useful to illustrate the Polyakov blocks.
		
		\begin{figure}[h]
			\center
			\begin{tikzpicture}
				\def\x{0.8}
				\filldraw[blue!10!white] (0,0) circle (4*\x cm);
				\draw[] (0,0) circle (4*\x cm);
				
				\draw[] (-2.8*\x,-2.8*\x) to (-1.5*\x,0);
				\draw[] (-2.8*\x,2.8*\x) to (-1.5*\x,0);
				\draw[] (2.8*\x,2.8*\x) to (1.5*\x,0);
				\draw[] (2.8*\x,-2.8*\x) to (1.5*\x,0);
				\draw[] (-1.5*\x,0) to (1.5*\x,0);
				
				\node[anchor =west] at (1.5*\x,0) {$-\lambda$};
				\node[anchor =east] at (-1.5*\x,0) {$-\lambda$};
				\node[anchor=south] at (0,0) {$\Delta_E$};
				
				\node[anchor=south east] at (-2.8*\x,2.8*\x) {$x_1$};
				\node[anchor=north east] at (-2.8*\x,-2.8*\x) {$x_4$};
				\node[anchor=north west] at (2.8*\x,-2.8*\x) {$x_3$};
				\node[anchor=south west] at (2.8*\x,2.8*\x) {$x_2$};
			\end{tikzpicture}
			\caption{Exchange Witten diagram for 4 external insertions of identical scalars of weight $\Delta_\phi$ exchanging a scalar of weight $\Delta_E$ in the $t$ channel}
			\label{Witten exchange}
		\end{figure}
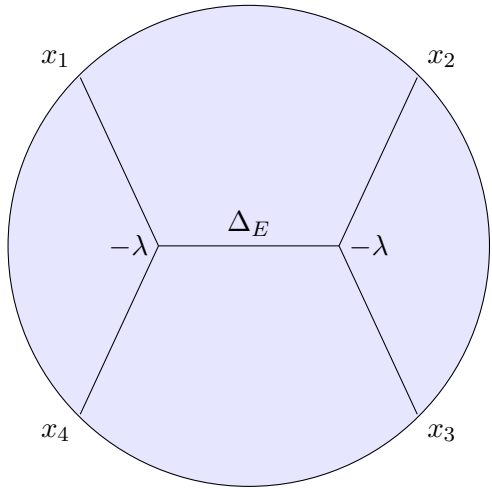
		
		The integral we are interested in, corresponding to the Witten diagram \eqref{Witten exchange}, is
		\begin{align}
			J^{(t)}(x_1,...,x_4) &= \int \frac{d^2w}{w_0^2}I(w,x_2,x_3)K_{\Delta_1}(x_1,w)K_{\Delta_4}(x_4,w),\\
			I(w,x_2,x_3) &= \int \frac{d^2z}{z_0^2}G_{\Delta_E}(w,z)K_{\Delta_2}(x_2,z)K_{\Delta_3}(x_3,z).\label{Eq: I three point}
		\end{align}
		The integral $I(z_a,x_a;x_2,x_3)$ is a boundary-boundary-to-bulk three-point function and has conformal symmetry. 
		
		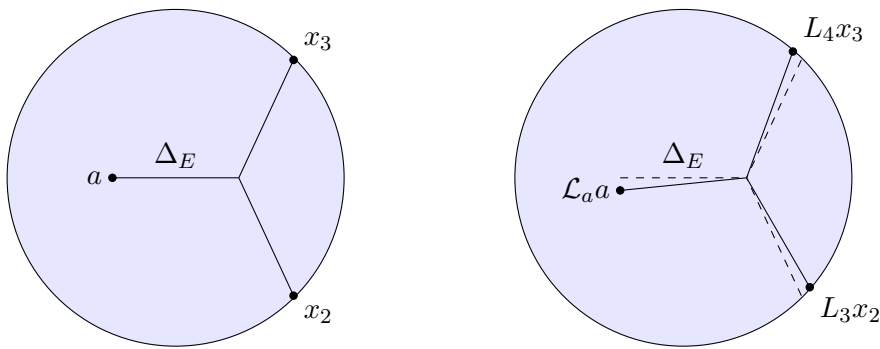
\begin{figure}[h!]
			\center
			\begin{tikzpicture}[scale=.7]
				\def\x{0.8}
				\filldraw[blue!10!white] (0,0) circle (4*\x cm);
				\draw[] (0,0) circle (4*\x cm);

				\draw[] (2.8*\x,2.8*\x) to (1.5*\x,0);
				\draw[] (2.8*\x,-2.8*\x) to (1.5*\x,0);
				\draw[] (-1.5*\x,0) to (1.5*\x,0);
				\filldraw[] (-1.5*\x,0) circle (2pt);
				\filldraw[] (2.8*\x,-2.8*\x)  circle (2pt);
				\filldraw[] (2.8*\x,2.8*\x)  circle (2pt);
				
				\node[anchor =east] at (-1.5*\x,0) {$a$};
				\node[anchor=south] at (0,0) {$\Delta_E$};
				
				\node[anchor=north west] at (2.8*\x,-2.8*\x) {$x_2$};
				\node[anchor=south west] at (2.8*\x,2.8*\x) {$x_3$};
			\end{tikzpicture}
			\hspace{2cm}
			\begin{tikzpicture}[scale=.7]
				\def\x{0.8}
				\filldraw[blue!10!white] (0,0) circle (4*\x cm);
				\draw[] (0,0) circle (4*\x cm);

				\draw[] (2.6*\x,3*\x) to (1.5*\x,0);
				\draw[] (3*\x,-2.6*\x) to (1.5*\x,0);
				\draw[] (-1.5*\x,-.3*\x) to (1.5*\x,0);
				\filldraw[] (-1.5*\x,-.3*\x) circle (2pt);
				\filldraw[] (3*\x,-2.6*\x)  circle (2pt);
				\filldraw[] (2.6*\x,3*\x)  circle (2pt);
				
				\node[anchor =east] at (-1.5*\x,-.3*\x) {$\mathcal{L}_a a$};
				\node[anchor=south] at (0,0) {$\Delta_E$};
				
				\node[anchor=north west] at (3*\x,-2.6*\x) {$L_3 x_2$};
				\node[anchor=south west] at (2.6*\x,3*\x) {$L_4 x_3$};

				\draw[dashed] (2.8*\x,2.8*\x) to (1.5*\x,0);
				\draw[dashed] (2.8*\x,-2.8*\x) to (1.5*\x,0);
				\draw[dashed] (-1.5*\x,0) to (1.5*\x,0);
			\end{tikzpicture}
			\caption{The system is invariant under an infinitesimal transformation \eqref{Eq: Infinitesimal transformation}.}
		\end{figure}
	
	\par 
		$I(w,x_2,x_3)$ in \eqref{Eq: I three point} is invariant under global transformations generated by $\mathbf{\mathcal{L}}_a+\vec{L}_2+\vec{L}_3$,\footnote{We write the transformations under the conformal group as vectors $\vec{L}_2 = (L_{-1},L_0,L_1)$ acting on point 2.} where the first term generates the isometries of $AdS_2$ and the other two generate the conformal transformation of the boundary. 
		As such, we can write
		\begin{align}
			(\mathbf{\mathcal{L}}_a+\vec{L}_2+\vec{L}_3)I(z_a,x_a;x_2,x_3) = 0, \label{Eq: Infinitesimal transformation}
		\end{align}
		and can therefore relate the Casimirs of the generators
		\begin{align}
			-\frac{1}{2}\mathbf{\mathcal{L}}_a^2 I (z_a,x_a;x_2,x_3) &=-\frac{1}{2} (\vec{L}_2+\vec{L}_3)^2I(z_a,x_a;x_2,x_3)\\
			&=C^{(23)}_2 I(z_a,x_a;x_2,x_3) .
		\end{align}
		In one dimension, the quadratic Casimir of the $AdS_2$ isometries is the Laplacian of the bulk \ref{App: quad casimir}. This Laplacian will allow us to get rid of the bulk-to-bulk propagator through the equation of motion \eqref{Eq: Equation of motion ads2}. Linking the previous elements together, we obtain
		\begin{align}
			(C_{(23)}^{(2)}-m^2_E)I(z_a,x_a;x_2,x_3) &= (\nabla_{a}^2-m_E^2)I(z_a,x_a;x_2,x_3)\\
			&= K_{\Delta_\phi}(z_a,x_a;x_2) K_{\Delta_\phi}(z_a,x_a;x_3).
		\end{align}
		The quadratic Casimir acting on the points 3 and 4 commutes with the other coordinates, so we can write a differential equation relating the full exchange diagram to the contact term
		\begin{align}
			(C_{(23)}^{(2)}-m^2_E)	J(x_1,x_2,x_2,x_3) &=\int \frac{dz_a dx_a}{z_a^2} \Pi_{i=1}^4K_{\Delta_\phi}(z_a,x_a;x_i)\\
			&= \frac{A}{(x_{13}x_{24})^{2\Delta_\phi}} \bar{D}_{\Delta_\phi \Delta_\phi \Delta_\phi \Delta_\phi}(\chi).
		\end{align}
		The same analysis holds for any two legs attached to a bulk-to-bulk propagator, though the final differential equation might depend on many variables.
		
		\subsection{Perturbative Polyakov blocks}\label{App: Polyakov}
		
		\subsection*{Polyakov blocks and OPE expansion}
		
		It will be useful to illustrate how Polyakov blocks and Conformal blocks operate perturbatively. 
		The four-point correlator of a scalar of conformal dimension $\Delta $ is
		\begin{align}
			\langle \phi_\Delta(x_1)\phi_\Delta(x_1)\phi_\Delta(x_1)\phi_\Delta(x_1)\rangle  = \frac{(C_\Delta)^4}{(x_{13}x_{24})^{2\Delta}}f(\chi)\\
			f(\chi) = \sum_h c^2_{\Delta\Delta h }\chi^{-2\Delta}G_h(\chi) = \sum_h c^2_{\Delta\Delta h }\chi^{-2\Delta}P_h(\chi).
		\end{align}
		where $G_h$ are the conformal blocks and $P_h(\chi)$ are the Polyakov blocks. The four-point conformal blocks in $d=1$ are the eigenfunctions of the quadratic Casimir
		\begin{align}\label{Eq: conformal blocks}
			G_h(\chi) = \chi^h {}_2F_1(h,h,2h;\chi).
		\end{align}
		The Polyakov blocks are not eigenvalues of the quadratic Casimir and depend non-trivially on the exchanged and external dimension. Though the Polyakov blocks are not known in closed form in position sapce, their double-discontinuity is equal to that of the conformal blocks in the $t$-channel
		\begin{align}
			dDisc[\chi^{-2\Delta}P^{(t)}_h(\chi)] &= dDisc[\chi^{-2\Delta}G^{(t)}_{h}(\chi)] \nonumber \\
			&= 2\sin^2\left(\frac{\pi}{2}(h-2\Delta)\right)(1-\chi)^{-2\Delta}G_h(1-\chi).
		\end{align}
		The double-discontinuity of the Polyakov block in the $t$-channel is given by the replacement $\chi\rightarrow 1-\chi$ since the Polyakov block is crossing-symmetric\footnote{We choose a prefactor $\frac{1}{(x_{13}x_{24})^{2\Delta}}$ to have a crossing-symmetric function $f(\chi)$. However, we keep the normalisation of the blocks in equation \eqref{Eq: conformal blocks} to be consistent with the literature and use the combination $\chi^{-2\Delta}P(\chi)$ to work with a truly crossing-symmetric quantity, without the need for a prefactor. }.
		\begin{align}
			dDisc[\chi^{-2\Delta}P^{(s)}_h(\chi)] = 2\sin^2\left(\frac{\pi}{2}(h-2\Delta)\right)\chi^{-2\Delta}G_h(\chi).
		\end{align}
		If we expand the four-point correlator in a small strong coupling parameter $\epsilon$
		\begin{align}
			f(\chi) = f^{0}(\chi)+\epsilon f^{1}(\chi)+O(\epsilon^2),
		\end{align}
		one can look at the structure and properties of these two expansions. The first order is Generalised Free Field Theory where the spectrum is $\{0,h=2\Delta+2n\}$, and the correlator is  obtained with the pairwise Wick contractions between fields
		\begin{align}
			f^{0}(\chi) =1+  \chi^{2\Delta}+\left(\frac{\chi}{1-\chi}\right)^{2\Delta}.
		\end{align}
		The conformal decomposition 
		\begin{align}
			f^0(\chi)&=1+\sum_{n}c_{2\Delta+2n,\Delta,\Delta}^2G_{2\Delta+2n}(\chi),
		\end{align}
		gives the OPE coefficients.
		Whereas the Polyakov blocks vanish at the position of the double trace operators
		\begin{align}
			f^0(\chi) = P_{\Delta,0}(\chi)+\sum_{n}c_n^{(0)}P_{2\Delta+2n}(\chi),
		\end{align}
		giving the identity contribution in all channels.
		\begin{align}
			P_{\Delta,0}(\chi) = 1+  \chi^{2\Delta}+\left(\frac{\chi}{1-\chi}\right)^{2\Delta}.
		\end{align}
		
		At first order in a perturbative expansion, assuming that there are no new exchanged operators\footnote{At strong coupling, this means that the first order Witten diagrams are contact diagrams and not exchange diagrams.}, the spectrum is $h=\{0,2\Delta+2n+\epsilon \gamma^{(1)}_n\}$, where the identity operator receives no corrections. The first order OPE expansion is then
		\begin{align}
			f^{1}(\chi) & = \sum_n c^{(1)}_nG_n(\chi)+c_n^{(0)}\left(\frac{g\gamma_n}{2}\right)\partial_nG_n(\chi)\\
			&= \sum_n c_n^{(0)}\left(\frac{g\gamma_n}{2}\right)\partial_nP_n(\chi).
		\end{align}
		Only one term in the Polyakov block expansion is non vanishing in this case
		\begin{align}
			f^{1}(\chi) = c_0^{(0)}\left(\frac{g\gamma_0}{2}\right)(\partial_nP_n)|_{n=0}(\chi).\label{Eq: contact Witten diagram}
		\end{align}
		If there is a new exchanged operator $\Delta_E\neq 2\Delta+2n$ at this order ($c_{\Delta_E\Delta\Delta}=O(\sqrt{\epsilon})$), the expansion is changed by a factor:
		\begin{align}\label{Eq: Polyakov pert exp}
			f^{1}(\chi) =c_0^{(0)}\left(\frac{g\gamma_0}{2}\right)(\partial_nP_n)|_{n=0}(\chi)+c_{\Delta_E\Delta\Delta}^2P_{\Delta_E}(\chi)
		\end{align}
		In the strong coupling language, this corresponds to having an Exchange Witten diagram with exchange dimension $\Delta_E$. Hence, the Polyakov blocks are given by exchange Witten diagrams, up to the contact diagram contribution from equation \eqref{Eq: contact Witten diagram}.
		
		\subsection*{Polyakov blocks and exchange Witten diagrams}
		The computation of the exchange Witten diagram in the main text \eqref{Eq: Exchange solution Delta=1} is the sum of solutions to second-order differential equations. The integration constants are provided by  the three-point function and the symmetry of the correlator. The example given in the main text corresponding to the Polyakov block of external dimension $\Delta=1$ and exchanged dimension $\Delta_E=1$ can be computed with the toy model of a massless scalar theory in AdS$_2$ with a $\phi^3$ interaction, this corresponds to the action
		\begin{align}
			S_{\phi^3} = \int \frac{dtdz}{z^2}\left( \partial_\mu \phi \partial^\mu \phi-\frac{\lambda}{3!} \phi^3\right).
		\end{align}
		The leading order is given by GFF. The next to leading order ($O(\lambda)$) correlators come from the constant vertex giving the 3 point function 
		\begin{align}
			<\phi(x_1)\phi(x_2)\phi(x_3)> & =\frac{\lambda}{\pi^3} \frac{\pi^2}{2x_{12}x_{23}x_{13}},
		\end{align}
		and the OPE coefficient
		\begin{align}
			c_{111}&= \frac{\lambda}{2\pi}.
		\end{align}
		The first sub-leading term ($O(\lambda^2)$) in the four-point function is generated by the exchange diagram which contributes
		\begin{align}
			<\phi(x_1)\phi(x_2)\phi(x_3)\phi(x_4)> &=\sum_{s,t,u} \lambda^2J_{\Delta_E=1,\Delta=1}(x_1,...,x_4)\\
			&=\frac{\lambda^2}{\pi^5}\frac{\pi}{4}\frac{1}{(x_{13}x_{24})^2}P^{(0)}_{(1,1)}(\chi),
		\end{align}
		where $P^{(0)}_{(1,1)}(\chi)$ is written explicitly in Equation \eqref{Eq: Exchange solution Delta=1} and has a small $\chi$ expansion\footnote{We have kept the integration constants from \eqref{Eq: fs1}.}
		\begin{align}
			<\phi(x_1)\phi(x_2)\phi(x_3)\phi(x_4)>&=\frac{1}{(x_{13}x_{24})^2} \frac{\lambda^2}{4\pi^4}\left(\frac{c_1-6 \zeta (3)}{\chi^2}+\frac{2 \left(\pi ^2-3 c_2 \right)}{3 \chi}\right)+o\left(\frac{1}{\chi}\right).
		\end{align}
		The first term in the expansion is set to zero since it corresponds to a correction to the identity operator. The second term corresponds in the conformal s-channel OPE to
		\begin{align}
			\frac{1}{(x_{13}x_{24})^2}\sum_h c_{11h}^2 \chi^{h-2} {}_2F_1(h,h,2h,\chi) & = \frac{1}{(x_{13}x_{24})^2}\left(\frac{c_{111}^2}{\chi}+o\left(\frac{1}{\chi}\right) \right),
		\end{align}
		where by equating the expansions in the conformal and Polyakov blocks, this corresponds to $c_{\Delta_E \Delta\Delta}$ in equation \eqref{Eq: Polyakov pert exp}. 
		The known three-point function and the four-point OPE then give the solutions for the integration constants
		\begin{align}
			c_1 = 6\zeta(3)\\
			c_2= -\frac{\pi^2}{6},
		\end{align}
		and provide the correct numerical factor for the Polyakov block computed in the main text as well as the maximally symmetric and Regge-bounded function.

		\subsection{Higher weight exchange diagrams and bootstrap}
		The process done in the main text can be repeated for other exchanged weights, for example we have the s-channel exchanges
		\begin{align}
			f^{(s)}_{\Delta_E=2} (\chi) &=\frac{4 c_2-4 \chi \log ^2(\chi)+8 \chi \log (\chi)+2 \log (1-\chi) (2 \chi \log (\chi)-4)+\frac{4 \pi ^2}{3}}{\chi^3} \\
			&+\frac{2 (\chi-2)}{\chi^3} \left( c_1-c_2 \log (1-\chi)-\text{Li}_2(\chi) \log \left(\frac{1-\chi}{\chi^2}\right)-\text{Li}_3(1-\chi)-2 \text{Li}_3(\chi)\right. \nonumber \\
			&+\left.\log \left(\frac{\chi}{1-\chi}\right) \log (1-\chi) \log (\chi)\right)\nonumber \\
			f^{(s)}_{\Delta_E=3} &= -\frac{2 ((\chi-6) \chi+6) (+3 \text{Li}_3(1-\chi)+\log (1-\chi) (\text{Li}_2(\chi)+\log (1-\chi) \log (\chi)))}{\chi^4}\\
			&-\frac{6 (\chi-2) (2 \text{Li}_2(\chi)+\log (1-\chi) \log (\chi))}{\chi^3}\nonumber\\
			&+\frac{2 \left(9 \chi^2 \log (\chi)+\left(\pi ^2 ((\chi-6) \chi+6)-9 \left(\chi^2+\chi-3\right)\right) \log (1-\chi)\right)}{3 \chi^4}\nonumber \\
			&+2 \frac{\left(\pi ^2 (\chi-3)-9 (\chi-4)\right) (\chi-1)}{\chi^4}\nonumber \\
			f^{(s)}_{\Delta_E=4} &=-\frac{2 (\chi-2) \left(\chi^2-10 \chi+10\right)}{\chi^5} \Big( \text{Li}_2(\chi) \log \left(\frac{1-\chi}{\chi^2}\right)\\
			&\qquad +\text{Li}_3(1-\chi)+2 \text{Li}_3(\chi)+\log (\chi) \log ^2(1-\chi)-\log ^2(\chi) \log (1-\chi)\Big)\nonumber \\
			&+\frac{2 \left(11 \chi^2-60 \chi+60\right) (\log (1-\chi)-\log (\chi)) \log (\chi)}{3 \chi^4}\nonumber\\
			&+\frac{8 \chi (\chi (19 \chi-90)+90) \log (\chi)-2 (\chi-1) (\chi (76 \chi-335)+310) \log (1-\chi)}{9 \chi^5}\nonumber\\
			&+\frac{10 (\chi (2 (104-9 \chi) \chi-505)+330)+2 \pi ^2 (\chi ((336-25 \chi) \chi-885)+610)}{9 \chi^5}\nonumber 
		\end{align}
		As we increase the weight of the exchanged operator, this seemingly increases the complexity of the solution, however, there are some patterns that are easy to spot. Notice that all transcendentality 3 terms have the same polynomial function multiplying them motivating the use a transcendentality ansatz \cite{Ferrero:2019luz}:
		\begin{align}
			\sum_{n,i} r_{i,n}(\chi) T_{i_n}(\chi)
		\end{align}
		And inserting it in the differential equations, we obtain differential equations for  the polynomials multiplying the transcendentality $i$ functions for arbitrary weight $\Delta_E$ and arbitrary external weights $\Delta$. 
		In particular, the polynomials multiplying the transcendentality 3 terms remaining after the action of the Casimir differential operator will satisfy the homogeneous differential equation. Explicitly we have
		\begin{align}
			r_{3,n}^{(s)} &= c_n  \chi^{-2 \Delta -\Delta_E+1}\, _2F_1(1-\Delta_E,1-\Delta_E;2-2 \Delta_E;\chi) \\
			r_{3,n}^{(t)} &= c_n  (1-\chi)^{-2 \Delta -\Delta_E+1}\, _2F_1(1-\Delta_E,1-\Delta_E;2-2 \Delta_E;1-\chi) \\
			r_{3,n}^{(u)}&= P_{\Delta_E-1}(2 \chi-1)
		\end{align}
		The transcendentality 2 functions are multiplied by polynomials that satisfy inhomogeneous differential equations and are difficult to solve for general $\Delta_E$. However, for specific cases, they are simple to solve. 
		We can do this for example for higher exchanged weights and more importantly, for higher-point correlators with exchanges. This analysis is beyond the scope of this paper and left for further investigation

		\section{Multipoint Ward identity: a check}
			\label{subsec: Ward Id}
\begin{figure}[h]
		\centering
		\begin{tikzpicture}[scale=.4]
			\def\x{4}
			\filldraw[blue!10!white] (0,0) circle (\x cm);
			\draw[] (0,0) circle (\x cm);
			
			\draw[] (-.5*\x, 0.866*\x)to (.5*\x, 0.866*\x);
			\node[anchor=south west] at (.5*\x, 0.866*\x){$x_6$};
			\node[anchor=south east] at (-.5*\x, 0.866*\x){$x_1$};
			
			\draw[] (-1*\x, 0*\x) to (\x, 0*\x);
			\node[anchor= west] at (\x, 0*\x){$x_5$};
			\node[anchor=south east] at (-1*\x, 0*\x){$x_2$};

			\draw[] (.5*\x, -0.866*\x) to (-.5*\x, -0.866*\x);
			\node[anchor= north east] at (-.5*\x, -0.866*\x){$x_3$};
			\node[anchor=north west] at (.5*\x, -0.866*\x){$x_4$};
			
		\end{tikzpicture} \hspace{.5cm}
		\begin{tikzpicture}[scale=.4]
			\def\x{4}
			\filldraw[blue!10!white] (0,0) circle (\x cm);
			\draw[] (0,0) circle (\x cm);
			
			\node[anchor=south west] at (.5*\x, 0.866*\x){$x_6$};
			\node[anchor=south east] at (-.5*\x, 0.866*\x){$x_1$};
			\node[anchor= west] at (\x, 0*\x){$x_5$};
			\node[anchor=south east] at (-1*\x, 0*\x){$x_2$};
			\node[anchor= north east] at (-.5*\x, -0.866*\x){$x_3$};
			\node[anchor=north west] at (.5*\x, -0.866*\x){$x_4$};
			
			\draw[] (-.5*\x, 0.866*\x)to  (\x, 0*\x);
			\draw[] (.5*\x, -0.866*\x) to (-.5*\x, -0.866*\x);
			\draw[] (-1*\x, 0*\x) to (-0.03*\x, 0.56*\x);
			\draw[] (0.03*\x, 0.59*\x) to (.5*\x, 0.866*\x);
		\end{tikzpicture}  \hspace{.5cm}
		\begin{tikzpicture}[scale=.4]
			\def\x{4}
			\filldraw[blue!10!white] (0,0) circle (\x cm);
			\draw[] (0,0) circle (\x cm);
			
			\node[anchor=south west] at (.5*\x, 0.866*\x){$x_6$};
			\node[anchor=south east] at (-.5*\x, 0.866*\x){$x_1$};
			\node[anchor= west] at (\x, 0*\x){$x_5$};
			\node[anchor=south east] at (-1*\x, 0*\x){$x_2$};
			\node[anchor= north east] at (-.5*\x, -0.866*\x){$x_3$};
			\node[anchor=north west] at (.5*\x, -0.866*\x){$x_4$};
			
			\draw[] (-.5*\x, 0.866*\x)to  (-.5*\x, 0.05*\x);
			\draw[] (-.5*\x, -0.05*\x)to  (-.5*\x, -0.866*\x);
			\draw[] (-1*\x, 0*\x) to (\x, 0*\x);
			\draw[] (.5*\x, -0.866*\x) to (.5*\x, -.05*\x);
			\draw[] (.5*\x, 0.05*\x) to (.5*\x, 0.866*\x);
		\end{tikzpicture} 
		\caption{A non-exhaustive list of leading order disconnected Witten diagram. The overlapping lines are broken to indicate they are not vertices, it is not a non-planar contribution at strong coupling, since we are in the strict planar limit.}
		\label{Fig: GFF witten 6pt}
	\end{figure}
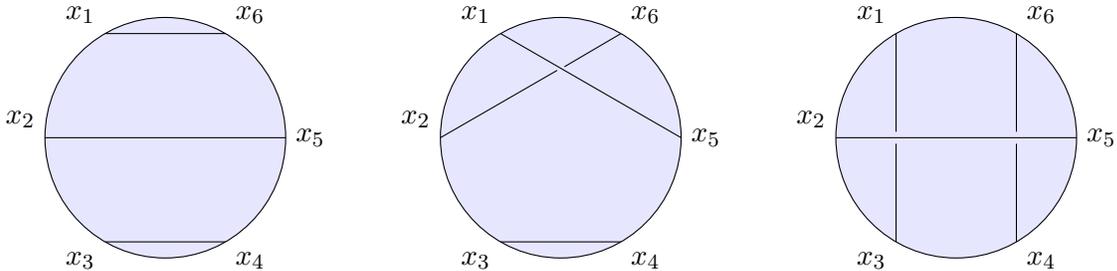

	A concrete example of $n$-point correlators of fields at the boundary of AdS$_2$ is the dual setup of the defect CFT of operator insertions on the 1/2 BPS Wilson line in $\mathcal{N}=4$ SYM studied at strong coupling in \cite{Giombi:2017cqn,Liendo:2018ukf}. In higher-point correlators, the first terms in the strong coupling expansion will originate from disconnected Witten diagrams. These first terms in the strong coupling expansion should satisfy the multipoint Ward identity presented in \cite{Barrat:2021tpn}, therefore providing a perturbative check of these conjectured Ward identities
	\begin{align}\label{Eq: WardId}
		\sum_{k=1}^{n-3} \left(\frac{1}{2}\partial_{\chi_k}+\alpha_k\partial_{r_k}-(1-\alpha_k)\partial_{s_k} \right)\mathcal{A}_{\Delta_1...\Delta_n}\raisebox{-.2cm}{$\Biggr|$}_{\raisebox{+.2cm}{$\begin{subarray}{l}
					r_i\rightarrow \alpha_i \chi_i \\s_i\rightarrow (1-\alpha_i)(1-\chi_i)\\t_{ij}\rightarrow (\alpha_i-\alpha_j)(\chi_i-\chi_j)
				\end{subarray}$}} =  0
	\end{align}\par 
	where we use the notation of \cite{Barrat:2021tpn} in which the cross-ratios are defined as\footnote{The shift is the index of the cross-ratios is to have them ranging from $1$ to $n-3$.}
	\begin{align}
		\chi_{i-1} &=\frac{x_{1i}x_{n-1,n}}{x_{in}x_{1,n-1}}\\
		r_{i-1}&=\frac{(u_{1}\cdot u_{i})(u_{n-1}\cdot u_{n})}{(u_{i}\cdot u_{n} )(u_{1}\cdot u_{n-1})}\\
		s_{i-1}&=\frac{(u_{1}\cdot u_{n})(u_{i}\cdot u_{n-1})}{(u_{i}\cdot u_{n} )(u_{1}\cdot u_{n-1})}\\
		t_{i-1,j-1} &=\frac{(u_{i}\cdot u_{j})(u_{1}\cdot u_{n})(u_{n-1}\cdot u_{n})}{(u_{i}\cdot u_{n} )(u_{j}\cdot u_{n} )(u_{1}\cdot u_{n-1})},
	\end{align}
	and the Ward identity is applied to the correlator of dimension $\Delta=1$, SO(5) vectors  expressed in terms of cross-ratios 
	\begin{align}
		\langle\Pi_{i=1}^{6}u_{a_i} \Phi^{a_i}(x_i)\rangle = C(x_i,u_i)\mathcal{A}_{\Delta=1,n=6}(\chi_{i},r_{i},s_i,t_{i,j})
	\end{align}
where the prefactor 
\begin{align}
	C(x_i,u_i)= \frac{(u_1.u_5)^2 (u_2.u_6) (u_3.u_6 )(u_4.u_6) }{(u_1.u_6) (u_5.u_6)}\frac{x_{16}^2 x_{56}^2}{ x_{15}^4 x_{26}^2 x_{36}^2 x_{46}^2},
\end{align}
has been amputated.
	The strong coupling expansion of this quantity is given by the holographic dual  studied in \cite{Giombi:2017cqn}, where the dual of these particular fields are the fluctuations in the S$^5$ directions propagating in the AdS$_2$ minimal string surface. 
	If we denote the strong coupling expansion parameter by $g$, we can expand the six-point function at strong coupling
	\begin{align}
		\mathcal{A}_{\Delta=1,n=6}(\chi_{i},r_{i},s_i,t_{i,j}) & = \mathcal{A}^{(0)}_{\Delta=1,n=6}(\chi_{i},r_{i},s_i,t_{i,j})+g \mathcal{A}^{(1)}_{\Delta=1,n=6}(\chi_{i},r_{i},s_i,t_{i,j})+O(g^2)
	\end{align} 
	Given the Ward identity is coupling independent, each of the terms in this perturbative expansion should satisfy equation \eqref{Eq: WardId} .
	The leading term, $\mathcal{A}^{(0)}_{\Delta=1,n=6}$, in the strong coupling expansion corresponds to the simple Wick contractions where a few examples are drawn in \ref{Fig: GFF witten 6pt}. 
\par
\begin{figure}
		\centering
		\begin{tikzpicture}[scale=.4]
			\def\x{4}
			\filldraw[blue!10!white] (0,0) circle (\x cm);
			\draw[] (0,0) circle (\x cm);
			
			\node[anchor=south west] at (.5*\x, 0.866*\x){$x_6$};
			\node[anchor=south east] at (-.5*\x, 0.866*\x){$x_1$};
			\node[anchor= west] at (\x, 0*\x){$x_5$};
			\node[anchor=south east] at (-1*\x, 0*\x){$x_2$};
			\node[anchor= north east] at (-.5*\x, -0.866*\x){$x_3$};
			\node[anchor=north west] at (.5*\x, -0.866*\x){$x_4$};

			\draw[] (-.5*\x, 0.866*\x) to (0*\x, 0.5*\x);
			\draw[] (.5*\x, 0.866*\x) to (0*\x, 0.5*\x);
			\draw[] (-.5*\x, -0.866*\x) to (0*\x, 0.5*\x);
			\draw[] (.5*\x, -0.866*\x) to (0*\x, 0.5*\x);
			\filldraw[]  (0*\x, 0.5*\x) circle (0.7*\x pt);
			\draw[] (-1*\x, 0*\x) to (-.22*\x, 0*\x);
			\draw[]   (-.15*\x, 0*\x) to (.15*\x, 0*\x);
			\draw[] (.22*\x, 0*\x) to (1*\x, 0*\x);
		\end{tikzpicture} \hspace{.5cm}
		\begin{tikzpicture}[scale=.4]
			\def\x{4}
			\filldraw[blue!10!white] (0,0) circle (\x cm);
			\draw[] (0,0) circle (\x cm);
			
			\node[anchor=south west] at (.5*\x, 0.866*\x){$x_6$};
			\node[anchor=south east] at (-.5*\x, 0.866*\x){$x_1$};
			\node[anchor= west] at (\x, 0*\x){$x_5$};
			\node[anchor=south east] at (-1*\x, 0*\x){$x_2$};
			\node[anchor= north east] at (-.5*\x, -0.866*\x){$x_3$};
			\node[anchor=north west] at (.5*\x, -0.866*\x){$x_4$};
			
			\draw[] (-.5*\x, 0.866*\x)to  (\x, 0*\x);
			\draw[] (.5*\x, -0.866*\x) to (-.5*\x, -0.866*\x);
			\draw[] (-1*\x, 0*\x) to (-0.03*\x, 0.56*\x);
			\draw[] (0.03*\x, 0.59*\x) to (.5*\x, 0.866*\x);
			\filldraw[]  (0,0.575*\x) circle (0.7*\x pt);
		\end{tikzpicture}  \hspace{.5cm}
		\begin{tikzpicture}[scale=.4]
			\def\x{4}
			\filldraw[blue!10!white] (0,0) circle (\x cm);
			\draw[] (0,0) circle (\x cm);
			
			\node[anchor=south west] at (.5*\x, 0.866*\x){$x_6$};
			\node[anchor=south east] at (-.5*\x, 0.866*\x){$x_1$};
			\node[anchor= west] at (\x, 0*\x){$x_5$};
			\node[anchor=south east] at (-1*\x, 0*\x){$x_2$};
			\node[anchor= north east] at (-.5*\x, -0.866*\x){$x_3$};
			\node[anchor=north west] at (.5*\x, -0.866*\x){$x_4$};
			
			\draw[] (-.5*\x, 0.866*\x)to  (-.5*\x, 0.05*\x);
			\draw[] (-.5*\x, -0.05*\x)to  (-.5*\x, -0.866*\x);
			\draw[] (-1*\x, 0*\x) to (\x, 0*\x);
			\draw[] (.5*\x, -0.866*\x) to (.5*\x,  0.866*\x);
			\filldraw[]  (0.5*\x,0) circle (0.7*\x pt);
		\end{tikzpicture} 
		\caption{Some of the next-to-leading order disconnected Witten diagrams}
		\label{Fig: Disconnected first order contribution}
	\end{figure}
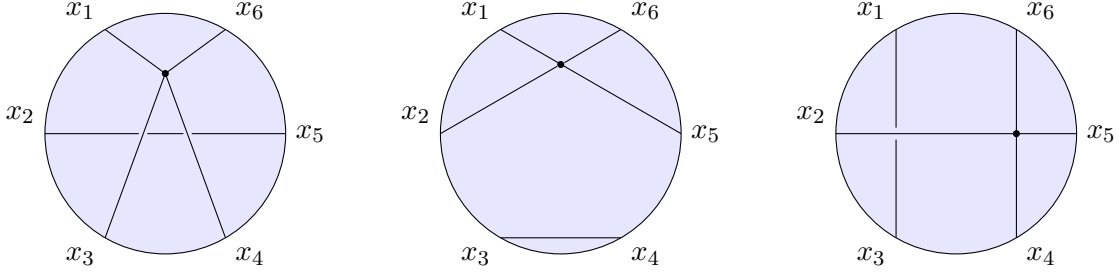

	Explicitly, performing these permutations, we find
	\begin{align}\label{Eq: A6 GFF}
		\mathcal{A}^{(0)}_{\Delta=1,n=6} &= \frac{r_3 t_{1,2}}{\left(\chi _1-\chi _2\right){}^2 \chi _3^2}+\frac{r_2 t_{1,3}}{\chi _2^2 \left(\chi _1-\chi _3\right){}^2}+\frac{r_1 t_{2,3}}{\chi _1^2 \left(\chi _2-\chi _3\right){}^2}+\frac{s_1 t_{2,3}}{\left(\chi _1-1\right){}^2 \left(\chi _2-\chi _3\right){}^2}\nonumber \\
		&+\frac{s_3 t_{1,2}}{\left(\chi _1-\chi _2\right){}^2 \left(\chi _3-1\right){}^2}+\frac{s_2 t_{1,3}}{\left(\chi _2-1\right){}^2 \left(\chi _1-\chi _3\right){}^2}+\frac{t_{1,2}}{\left(\chi _1-\chi _2\right){}^2}+\frac{t_{1,3}}{\left(\chi _1-\chi _3\right){}^2}\nonumber \\
		&+\frac{t_{2,3}}{\left(\chi _2-\chi _3\right){}^2}+\frac{r_2 s_1}{\left(\chi _1-1\right){}^2 \chi _2^2}+\frac{r_3 s_1}{\left(\chi _1-1\right){}^2 \chi _3^2}+\frac{r_1 s_2}{\chi _1^2 \left(\chi _2-1\right){}^2}+\frac{r_1 s_3}{\chi _1^2 \left(\chi _3-1\right){}^2}\nonumber \\
		&+\frac{r_2 s_3}{\chi _2^2 \left(\chi _3-1\right){}^2}+\frac{r_3 s_2}{\left(\chi _2-1\right){}^2 \chi _3^2}
	\end{align}
	For which, not only is the Ward identity satisfied, but each of the terms in equation \eqref{Eq: A6 GFF} corresponding to the different R-symmetry channels individually satisfy this equation. As such, this is a trivial check of this Ward identity.
	The next-to-leading term corresponds to the disconnected diagram with one Wick contraction and a contact four-point function,
	\begin{align}
		\mathcal{A}^{(1)}_{\Delta=1,n=6}(\chi_{i},r_{i},s_i,t_{i,j}) =\frac{r_1 t_{2,3} G^{(1)}_4\left(\frac{\chi _1 \left(\chi _2-\chi _3\right)}{\chi _2 \left(\chi _1-\chi _3\right)}\right)}{ \chi _1^2 \left(\chi _2-\chi _3\right){}^2}+\textrm{permutations }[(x_i,u_i)].
	\end{align}
	This corresponds to the permutations of which a few examples are given by the disconnected Witten diagrams in Figure \ref{Fig: Disconnected first order contribution}.
	The four-point function is computed in \cite{Giombi:2017cqn} and can be written as
	\begin{align}
		\hspace{-.5cm}&\langle \Phi^{a_1}(x_1)\Phi^{a_2}(x_2)\Phi^{a_3}(x_3)\Phi^{a_4}(x_4)\rangle = \frac{\delta^{a_1}_{a_2}\delta^{a_3}_{a_4}}{(x_{12}x_{34})^2}G_4^{(1)}(\frac{x_{12}x_{34}}{x_{13}x_{24}}) +\textrm{permutations }[(x_i,u_i)]\\
		&G_4^{(1)}(\chi)= \frac{(\chi -1) \left(\chi ^2+\chi +2\right) \log \left((\chi -1)^2\right)}{2 \chi }\!-\!\frac{\left(\chi ^2-2 \chi +2\right) \left(\chi ^2 \log \left(\chi ^2\right)-2 \chi +2\right)}{2 (\chi -1)^2}.
	\end{align}
	Just as in the GFF case, each of the R-symmetry channels also satisfy the Ward identities \eqref{Eq: WardId} individually. As a consequence, this is not constraining for the 6-point function, but rather motivates the consistency of these Ward identities. This seems to follow from the fact that the disconnected contribution can be written as a product of correlators, each satisfying a Ward identity, though this is dependent on the crossing and analytic properties of the four-point function. While it does not give insight into the connected contribution, this seems to indicate a recursive method to prove this Ward identity for any disconnected correlators.

	\bibliographystyle{nb}
	\bibliography{99-Ref_n_Witten.bib}
	
\end{document}